\documentclass[prb,twocolumn,showpacs,superscriptaddress]{revtex4-2}
\usepackage{epsfig}
\usepackage{amsmath,amssymb}
\usepackage{mathtools}
\usepackage{graphicx}
\usepackage{footnote}

\usepackage[normalem]{ulem}
\usepackage{color}

\definecolor{orange}{rgb}{1, 0.5, 0}

\definecolor{darkgreen}{rgb}{0, 0.5, 0}

\newcommand{\nuall}{\ensuremath{\nu_1\nu_2\nu_3\nu_4}}

\usepackage[all]{xy}

\renewcommand{\ddag}{d^\dagger}
\newcommand{\cdag}{c^\dagger}
\newcommand{\varepstilde}{\widetilde{\varepsilon}}
\newcommand{\dtau}[1]{\partial_{\tau_{#1}}}
\newcommand{\ceq}[1]{Eq.~(\ref{#1})}
\newcommand{\cfg}[1]{Fig.~\ref{#1}}

\usepackage{hyperref}
\begin{document}

\title{Symmetric Improved Estimators for Continuous-time Quantum Monte Carlo}

\author{J.~Kaufmann}
\thanks{These two authors contributed equally.}
\affiliation{\small\em Institute for Solid State Physics, TU Wien, 1040 Vienna, Austria}
\affiliation{\small\em Institute for Theoretical Solid State Physics, IFW Dresden, 01069 Dresden, Germany}

\author{P.~Gunacker}
\thanks{These two authors contributed equally.}
\affiliation{\small\em Institute for Solid State Physics, TU Wien, 1040 Vienna, Austria}

\author{A.~Kowalski}
\affiliation{\small\em Institut f\"ur Theoretische Physik und Astrophysik, Universit\"at W\"urzburg, 97074 W\"urzburg, Germany}

\author{G.~Sangiovanni}
\affiliation{\small\em Institut f\"ur Theoretische Physik und Astrophysik and W\"urzburg-Dresden Cluster of Excellence ct.qmat, Universit\"at W\"urzburg, 97074 W\"urzburg, Germany}

\author{K.~Held}
\affiliation{\small\em Institute for Solid State Physics, TU Wien, 1040 Vienna, Austria}

\date{\small\today}
\begin{abstract}
We derive equations of motion for Green's functions of the multi-orbital Anderson impurity model 
by differentiating symmetrically with respect to all time arguments. 
The resulting equations relate the one- and two-particle Green's function to correlators of up to six particles at four times.
As an application we consider continuous-time quantum Monte Carlo simulations in the hybridization expansion, 
which hitherto suffered from notoriously high noise levels at large Matsubara frequencies. 
Employing the derived symmetric improved estimators overcomes this problem.
\end{abstract}
\maketitle

\section{Introduction} \label{sec:Intro}
The Hubbard model (HM)\cite{Hubbard64} and the Anderson impurity model (AIM)\cite{Anderson}, which are related through the dynamical mean-field theory (DMFT)\cite{Georges1996}, are two of the basic models for strongly correlated electrons.
Thus, numerical and analytic solutions of these models over a wide range of parameters 
are of great interest in condensed matter physics.
As of today, the continuous-time quantum Monte Carlo (CT-QMC)\cite{Gull} method 
is the workhorse for obtaining numerical solutions in terms of one- and two-particle Green's functions.

CT-QMC algorithms are based on a stochastic sampling of the diagrammatic series
expansion of either the partition function or, directly, 
the thermal expectation value of some 
operators, which is also known as worm sampling.\cite{GullPhd,Gunacker}
For the AIM, one distinguishes between weak coupling expansions in the interaction CT-INT\cite{Rubstov_ct_int}
(and related CT-AUX\cite{Gull_aux}) and strong coupling expansions in the hybridization CT-HYB.\cite{Werner06} 
While traditionally the former are employed for single-orbital impurity model 
calculations and impurity clusters, the latter are primarily used for multi-orbital impurity models.
The reason for this is that in weak coupling, the exponential scaling of the sign problem makes multi-orbital calculations
with non-density-density interactions difficult, 
while in strong coupling, the exponential scaling of the local Hamiltonian dimensions
forbids medium- to large scale clusters.   

Conceptionally one might expect CT-INT and CT-HYB to behave similarly
apart from their differences in scaling with interaction and hybridization strength, respectively. 
The estimators for one- and two-particle Green's functions 
are instead considerably different in weak- and strong coupling approaches.
While Green's function estimators in CT-INT are formulated as corrections
to the non-interacting Green's function\cite{Gull_perfomance}, this is not the case for CT-HYB.
This results in poor asymptotic behavior of derived quantities, 
which in turn triggered a series of developments in the CT-HYB community attempting 
to remedy this problem. These developments include orthogonal polynomial 
representations as effective low-pass filters\cite{Boehnke}, moment 
expansions in the one-particle self-energy\cite{Potthoff,sigmatails}, asymptotic
expressions for the two-particle vertex functions\cite{Wentzell,Kaufmann} 
and approaches based on the equation of motion (EOM), often referred to as improved estimators\cite{Hafermann,Hafermann_ret,Gunacker2016a}.
Similar to CT-INT, the improved estimators of CT-HYB calculate the Green's function
as a correction to the non-interacting Green's function.

In this paper we introduce and explore the concept of symmetric improved estimators.
That is, we derive  EOMs by differentiating with respect to all time arguments. 
The derived equations relate the one- and two-particle Green's function to 
correlation functions of higher order in the number of creation and annihilation operators. 
These EOMs may prove useful in various contexts. We employ them for CT-HYB estimators 
of the self-energy, two-particle (four-leg) and three-leg vertex.
We give theoretical arguments showing that the symmetric improved estimators 
have a strongly reduced noise level at large Matsubara frequencies 
$\nu$, 
and even a different scaling with $\nu$, which is confirmed in actual CT-HYB calculations.

Section~\ref{sec:model} introduces the AIM Hamiltonian and our notation for the one- and two-particle Green's function. 
Section~\ref{sec:Theory} sketches the
derivations for the symmetric improved estimators on the one- and two-particle level. 
We note that the expressions are based on hierarchies of EOMs and are also 
useful outside the context of CT-QMC.\cite{Grzegorz,Moutenet}  We further
discuss the improved Monte Carlo error scaling of Green's functions, self-energies and vertex 
functions at large   Matsubara frequencies.
Section~\ref{sec:Implementation} discusses the implementation of the symmetric 
improved estimators in CT-HYB. An implementation for density-density interactions 
in segment CT-HYB\cite{Werner} is straight-forward, 
an implementation for general interactions requires worm sampling. 
We discuss drawbacks for a classical worm sampling implementation and propose methods to increase the sampling efficiency. 
In Section~\ref{sec:Validation} we show results for self-energies as well as three- and four-leg vertex functions 
and validate them by comparing to results from exact diagonalization (ED).
Finally, Section~\ref{sec:Conclusion} summarizes our work.
We furthermore provide an appendix that contains the derivation of the presented formulas in more detail. 

\section{Definitions and model}
\label{sec:model}
The AIM consists of an impurity site that is coupled to a bath and its Hamiltonian reads
\begin{eqnarray}
  \label{eq:aimham}%
H_{\mathrm{AIM}} &=& \frac{1}{2} \sum_{ijkl} U^{\phantom{\dagger}}_{ijkl} d^\dagger_{i} d^\dagger_{j} d^{\phantom{\dagger}}_{l} d^{\phantom{\dagger}}_{k} + \sum_{i} \tilde{\varepsilon}^{\phantom{\dagger}}_{i} d_{i}^\dagger d_{i}^{\phantom{\dagger}}+ \\
&&\!\!  +\! \sum_{K i}\!  \varepsilon^{\phantom{\dagger}}_{K i} c^\dagger_{Ki} c^{\phantom{\dagger}}_{Ki}
+\!  \sum_{K ij} \left[ V_{K}^{ij} c^\dagger_{K i} d^{\phantom{\dagger}}_{j} + ( V_{K}^{j i} )^* d^\dagger_i c^{\phantom{\dagger}}_{K j}\right]  . \nonumber
\end{eqnarray}
Here, $d_{i}$ ($d_{i}^\dagger$) is the annihilation (creation) operator of an electron with spin-orbital flavor $i$ on the impurity;
$c_{Ki}$ ($c_{Ki}^{\dagger}$) is the annihilation (creation) operator of an electron with impurity flavor $i$
in the non-interacting bath and $K$ subsumes the remaining bath degrees of freedom (e.~g.~the momentum $\mathbf{k}$).
The impurity is described by a local one-particle potential $\tilde{\varepsilon}_i$ (e.g. the crystal field),
the interaction matrix $U_{ijkl}$, the bath dispersion $\varepsilon_{Ki}$, and the hybridization strength $V_{K}^{ij}$.
Since the bath degrees of freedom appear in Eq.~\eqref{eq:aimham} at most quadratically, they can be formally integrated out yielding a one-body term in the impurity operators. This contains the hybridization function $\Delta_{ab}(\tau_1, \tau_2)$ encoding the entire influence of the bath.
In the following we assume a diagonal hybridization 
$\Delta_{ab}(\tau_1,\tau_2) \equiv \Delta_a(\tau_1,\tau_2)\delta_{ab}$
resulting in a diagonal one-particle Green's function $G^{\tau_1 \tau_2}_{ab} = G^{\tau_1 \tau_2}_{a} \delta_{ab}$. But the same concept of  symmetric improved estimator equations can be  extended straight-forwardly to non-diagonal hybridizations.

The hybridization function is most conveniently written in Matsubara frequencies as
\begin{equation}
  \label{eq:def-hyb-matsu}
  \Delta_a^\nu = \sum_K \frac{V_K^{aa}(V_K^{aa})^\ast}{i\nu - \varepsilon_{Ka}},
\end{equation}
and it relates to the non-interacting Green's function of the impurity as
\begin{equation}
  \label{eq:def-g0}
  \mathcal{G}_a^\nu = \frac{1}{i\nu - \tilde{\varepsilon}_a - \Delta_a^\nu} \; .
\end{equation}

We define the interacting one-particle Green's function of the AIM as
\begin{equation}
G_{a}(\tau_1, \tau_2) = -\langle T_\tau d^{\phantom{\dagger}}_a(\tau_1) d^\dagger_a(\tau_2)  \rangle, 
\end{equation}
where $d_a(\tau)$ ($d_a^\dagger(\tau)$) are now the annihilation (creation) operators 
for electrons of flavor $a$ at (imaginary) time $\tau$.
Furthermore, $T_\tau$ is the imaginary-time ordering operator, 
and $\langle \ldots \rangle = (\text{Tr} e^{-\beta H} \ldots)/Z $ the thermal expectation value
at temperature $T$ ($\beta=1/T$), $Z$ is the partition function.
The Green's function is related to the density as $n_a\equiv \langle d^\dagger_a d^{\phantom{\dag}}_a \rangle = 1+G_a(0+,0)$. 

Analogously, we define the two-particle Green's function of the AIM as
\begin{equation}
  G_{abcd}(\tau_1, \tau_2, \tau_3, \tau_4) = \langle T_\tau d^{\phantom{\dag}}_a(\tau_1) d^\dagger_b(\tau_2) d^{\phantom{\dag}}_c(\tau_3) d^\dagger_d(\tau_4) \rangle.
\end{equation}

The Fourier transforms to fermionic Matsubara frequencies\cite{Matsubara1955} $\nu = (2l + 1)\pi/\beta$ at integer numbers $l$ are given by
\begin{equation}
  G^\nu_{a} = \frac{1}{\beta} \int_0^\beta d\tau_1 d\tau_2 e^{i\nu(\tau_1 - \tau_2)} G_{a}(\tau_1, \tau_2)
\end{equation}
and 
\begin{multline}
  G^{\nu_1 \nu_2 \nu_3 \nu_4}_{abcd} = \frac{1}{\beta^2} \int_0^\beta  d\tau_1 d\tau_2 d\tau_3 d\tau_4 \\
  e^{i(\nu_1 \tau_1 - \nu_2 \tau_2 + \nu_3 \tau_3 - \nu_4 \tau_4)} G_{abcd}(\tau_1, \tau_2, \tau_3, \tau_4)
\end{multline}
for the one- and two-particle Green's function, respectively.

The time translation symmetry of $H_{\mathrm{AIM}}$  results in time translation symmetry of Green's functions in imaginary time.
This is equivalent to energy conservation, which for the two-particle Green's function reads
\begin{equation}
\label{eq:energy-conservation}
\nu_1 + \nu_3 = \nu_2 + \nu_4.
\end{equation}
For the two-particle Green's function it is more common to assume a mixed bosonic-fermionic frequency representation with two fermionic- and one bosonic frequency. 
However, the choice of these frequencies is ambiguous and therefore the derivations of this paper are done in four fermionic frequencies. 
With the definitions above, the reader can easily adapt the results to their favorite convention. (Sometimes we use generic bosonic frequencies $\omega$ which are however replaced by  $\nu_1\ldots \nu_4$ in the final expression.)

\section{Symmetric improved estimators} \label{sec:Theory}
\subsection{One-particle estimator}
\label{sec:1pEst}
It is well-known that the Heisenberg equation of motion for a one-particle Green's function
is an equation that connects the one- and two-particle Green's functions.
This has been exploited in the CT-HYB\cite{Hafermann,Hafermann_ret,Gunacker2016a} and numerical renormalization group (NRG)\cite{Bulla1998} algorithms. It leads to the 
so-called ``improved estimator'' equation
\begin{equation}
  \label{eq:ie-1p}
  G_a^\nu = \mathcal{G}_a^\nu \big(1 + \xi_a^\nu\big),
\end{equation}
where 
\begin{equation}
  \xi_{ab}^\nu = \xi_a^\nu \delta_{ab} = \frac{1}{\beta} \int_0^\beta d\tau_1 d\tau_2  \underbrace{\langle -T_\tau q^{\phantom{\dag}}_a(\tau_1) d_b^\dag(\tau_2) \rangle}_{\equiv  \xi_a(\tau_1,\tau_2)} e^{i\nu(\tau_1-\tau_2)}
\end{equation}
is a two-particle Green's function and $q$ contracts three operators at equal time,\footnote{
One may also refer to $\xi$ as $G\Sigma$.}
i.e.,
\begin{align}
q_a &= \sum_{jkl} U_{[aj]kl}\ddag_j d_l d_k \; , \\
q^\dagger_{a} &= \sum_{mno} U_{mn[ao]}\ddag_m \ddag_n d_o.
\end{align}
The explicit derivation of this equation of motion can be found in Appendix \ref{app:1P_sym} and is based on the derivative with respect to the first time argument of $G_{a}$.

In this paper, we now express $\xi_a(\tau_1,\tau_2)$ by using the equation of motion again. This time, we  apply it to the second time argument of $\xi_a(\tau_1,\tau_2)$, insert it into \ceq{eq:ie-1p},
and finally arrive at (for the detailed calculation see Appendix \ref{app:1P_sym})
\begin{equation}
  \label{eq:sie-1p}
  G_a^\nu = \mathcal{G}_a^\nu \bigg( 1 + \mathcal{G}_a^\nu \Big( 2\sum_{j} U_{[aj][aj]}n_j + \vartheta_a^\nu \Big) \bigg),
\end{equation}
which is also shown by Feynman diagrams in \cfg{fig:dyson-sie-diagrams}.
Here, we make use of the anti-symmetric $U$-matrix Eq.~\eqref{eq:u-asy} in Appendix \ref{app:1P_sym}
\begin{align}
  \label{eq:u-asy_main}
&\frac{1}{2} \left( U_{ijkl} - U_{jikl} \right) \eqqcolon U_{[ij]kl} \\
 &\frac{1}{2} \left( U_{ijkl} - U_{ijlk} \right) \eqqcolon U_{ij[kl]}.
\end{align}
and the following three-particle Green's function with only two distinct
time arguments is employed
\begin{equation}
  \vartheta_{ab}^\nu = \vartheta_a^\nu \delta_{ab} 
  = -\frac{1}{\beta} \int_0^\beta d\tau_1 d\tau_2 \langle T_\tau q^{\phantom{\dag}}_a(\tau_1) q_b^\dag(\tau_2) \rangle e^{i \nu (\tau_1 - \tau_2)}.
\end{equation}
\begin{figure}
  \includegraphics[width=0.45\textwidth]{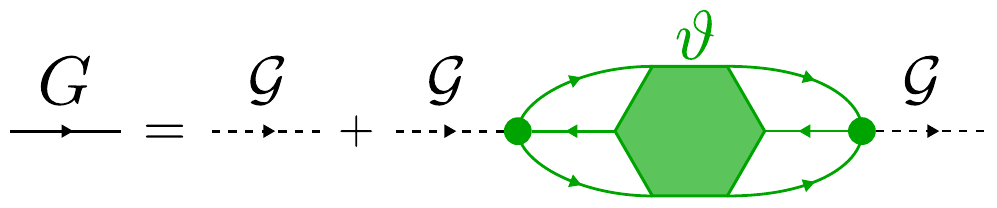}
  \caption{\label{fig:dyson-sie-diagrams} 
    Feynman-diagrammatic visualization of the symmetric improved estimator \ceq{eq:sie-1p} for the one-particle Green's function.
    The part of the diagram that is drawn in green, will be computed by CT-QMC in the following.
    Solid lines are full one-particle Green's functions $G$, dashed lines are non-interacting 
    Green's functions $\cal G$, and dots are $U$-matrices. The three-particle Green's function is
    represented by the hexagon. 
    Note that the Hartree-term $\propto\sum_j U_{[aj][aj]} n_j$ is excluded.}
\end{figure}

\paragraph*{Error propagation.}\label{sec:propagation}
The declared goal of improved estimators is to reduce high-frequency noise
in the quantity of interest. Among one-particle quantities, noise is
most prominent in the self-energy. Using \ceq{eq:sie-1p} and the Dyson equation,
we get
\begin{equation}
  \label{eq:se-sie}
  \Sigma_a^{\nu} = \frac{2\sum_j U_{[aj][aj]} n_j + \vartheta_a^\nu}{1 + \mathcal{G}_a^{\nu}(2\sum_j U_{[aj][aj]} n_j + \vartheta_{a}^\nu)}.
\end{equation}
The variance of the self-energy can hence approximately be computed by error propagation as (neglecting the error in the density $n_i$):
\begin{equation}
  \label{eq:var-se}
  \mathrm{var}[\Sigma] = \left|\frac{\partial\Sigma}{\partial\vartheta}\right|^2 \mathrm{var}[\vartheta].
\end{equation}
Since the derivative goes to 1 in the limit of high frequencies, we conclude that
the high-frequency noise amplitude is indeed identical to the noise amplitude of
the estimator $\vartheta_a^\nu$, which should be independent of $\nu$ for large $\nu$. This result for the symmetric improved estimator can be compared to the behavior of the conventional improved estimator for which  a discussion of the error propagation can be found elsewhere.\cite{Wallerberger16} The main result is that the error of the  conventional improved estimator grows linearly with $\nu$.

\subsection{Two-particle estimator}\label{sec:2pgf}
\begin{table*}
  \begin{tabular}{l l}
    Name \hspace{0.5cm} & Formula \\ \hline
    $\xi_{ab}^\nu $ & $ -\int_0^\beta d\tau_1 d\tau_2 \langle T_\tau q_a(\tau_1) d_b^\dag(\tau_2) \rangle e^{i\nu(\tau_1-\tau_2)}/\beta$\\
    $\vartheta_{ab}^\nu$ & $-\int_0^\beta d\tau_1 d\tau_2 \langle T_\tau q_a(\tau_1) q_b^\dag(\tau_2) \rangle e^{i \nu (\tau_1 - \tau_2)}/\beta$\\
    $\phi_{abcd}^{\omega}$ & $\int d\tau_1 d\tau_2 \langle T_\tau (Un)_{ab}(\tau_1) (Un)_{cd}(\tau_2) \rangle e^{i\omega(\tau_1-\tau_2)}/\beta$\\
    $\psi_{abcd}^{\omega}$ & $\int d\tau_1 d\tau_2 \langle T_\tau (Udd)_{ac}(\tau_1) (Ud^\dag d^\dag)_{bd}(\tau_2)\rangle e^{i\omega(\tau_1 - \tau_2)}/\beta$\\
    $f^{\nu\omega}_{abcd}$ & $\int d\tau_1 d\tau_2 d\tau_3 \langle T_\tau q_a(\tau_1) q_b^\dag(\tau_2) (Un)_{cd}(\tau_3) \rangle e^{i\nu(\tau_1\!-\!\tau_2)+i\omega(\tau_2\!-\!\tau_3)}/\beta$\\
    $g^{\nu\nu'}_{abcd}$ & $\int d\tau_1 d\tau_2 d\tau_3 \langle T_\tau q_a(\tau_1) q_c(\tau_2) (Ud^\dag d^\dag)_{bd}(\tau_3)\rangle e^{i\nu(\tau_1-\tau_3)+i\nu'(\tau_2-\tau_3)}/\beta$\\
    $h^{\nu_1\nu_2\nu_3\nu_4}_{abcd}$ & $\int d\tau_1 d\tau_2 d\tau_3 d\tau_4 \langle T_\tau q_a(\tau_1) q_b^\dag(\tau_2) q_c(\tau_3) q_d^\dag(\tau_4)\rangle e^{i(\nu_1\tau_1 - \nu_2\tau_2 + \nu_3\tau_3 - \nu_4\tau_4)}/\beta$
  \end{tabular}
  \caption{\label{tab:estimators}Terms of the symmetric improved estimators. We use the abbreviations $(Un)_{ab} = \sum_{jk}U_{[aj][bk]}d^\dag_j d_k$, 
    $(Ud^\dag d^\dag)_{ab} = \sum_{jk}U_{jk[ab]}d^\dag_j d^\dag_k$ and $(Udd)_{ab} = \sum_{jk}U_{[ab]jk}d_k d_j$. 
    Frequencies $\nu_{(i)}$ and $\omega$ are fermionic and bosonic Matsubara frequencies, respectively. }
\end{table*}
For the symmetric improved estimator of the two-particle Green's function, we obtain (again see the Appendix \ref{app:2p}
for the derivation):

\begin{widetext}
\begin{equation}
  \label{eq:2pgf-freq-1}
  G^{\nuall}_{abcd} = \mathcal{G}^{\nu_1}_a \Bigg(
    -R_{1,abcd}^{\nuall} + \mathcal{G}^{\nu_2}_b \bigg(
       R_{2,abcd}^{\nuall} + \mathcal{G}^{\nu_3}_c \Big(
        -R_{3,abcd}^{\nuall} + \mathcal{G}^{\nu_4}_d \big(
          R_{4,abcd}^{\nuall} + h_{abcd}^{\nuall}/\beta
        \big)
      \Big)
    \bigg)
  \Bigg) \mbox{ where }
\end{equation}
\begin{eqnarray}
  \label{eq:r1}
  R_{1}^{\nuall} &=& -\delta_{12}  G_d^{\nu_4}
  + \delta_{14}G_b^{\nu_2},\\
  \label{eq:r2}
  R_2^{\nuall} &=& -\delta_{12} \mathcal{G}_c^{\nu_3} 2\sum_{j}U_{[aj][aj]}n_j \!-\! \delta_{14} \xi_d^{\nu_4}\notag\\
  &&-\frac{2}{\beta}\mathcal{G}_c^{\nu_3} \mathcal{G}_d^{\nu_4} \bigg[\!-\!U_{[ac][bd]}
    +\!\! \sum_i\!\Big(U_{[ai][bd]} \xi_{ci}^{\nu_3} +\! U_{[ac][bi]}\xi_{di}^{\nu_4}\Big)
   \! -\! 2\phi_{abcd}^{\nu_4\!-\!\nu_3}\! +\! f_{cdab}^{\nu_3(\nu_3\!-\!\nu_4)}\!
  \bigg],\\
  \label{eq:r3}
  R_{3}^{\nuall} &=& \delta_{12} \vartheta_a^{\nu_1} + \frac{1}{\beta}\mathcal{G}_d^{\nu_4} \bigg[ 
    2\sum_i\Big(U_{[ic][bd]}\xi_{ai}^{\nu_1} + U_{[ac][id]}\xi_{bi}^{\nu_2}\Big)
    - \psi_{abcd}^{\nu_1+\nu_3} + 4 \phi_{adcb}^{\nu_1-\nu_4}
    + 2 f_{adcb}^{\nu_1(\nu_1-\nu_4)} - g_{dcba}^{\nu_4 \nu_2}\bigg],\\
  \label{eq:r4}
  R_{4}^{\nuall} &=& \frac{1}{\beta} \Big[2 f_{cbad}^{\nu_3(\nu_4-\nu_1)} + g_{abcd}^{\nu_1\nu_3} - 2 f_{abcd}^{\nu_1(\nu_1-\nu_2)}\Big].
\end{eqnarray}
\end{widetext}
\begin{figure}
  \includegraphics[width=0.8\textwidth]{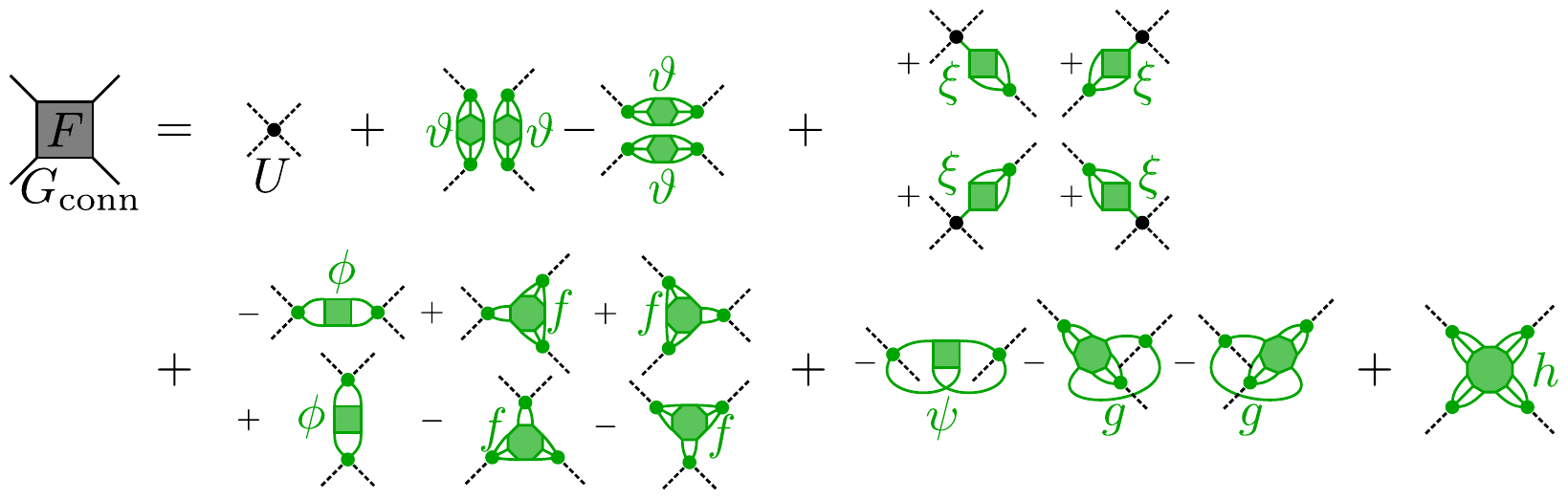}
  \caption{\label{fig:gconn-diagrams} 
    Heuristic drawing of the Feynman-diagrammatic
    decomposition of the connected part of the two-particle Green's function, 
    as obtained by symmetric improved estimators. The estimators that are later
    computed by CT-QMC are drawn in green. Solid lines represent interacting one-particle Green's functions $G$,
    dashed lines are non-interacting Green's functions $\mathcal{G}$, and dots are $U$-matrices.
    The terms involving $\vartheta$ are to be understood as products. 
    Note that Hartree-like terms and products thereof are not shown, in order to make the picture more concise.}
\end{figure}
All estimators occurring on the right-hand side of Eqs.~(\ref{eq:r1})-(\ref{eq:r4}) are defined in Table~\ref{tab:estimators}. 
In order to obtain a more symmetric form, we further made use of the relations
$\delta_{\nu_1\nu_2}\equiv\delta_{\nu_1\nu_2}\delta_{\nu_3\nu_4}$ and 
$\delta_{\nu_1\nu_4}\equiv\delta_{\nu_2\nu_3}\delta_{\nu_3\nu_4}$, which
are valid due to energy conservation. 
Additionally we employed
$\delta_{12}\equiv\delta_{ab} \delta_{cd} \delta_{\nu_1\nu_2}\delta_{\nu_3\nu_4}$
and $\delta_{14}\equiv\delta_{ad} \delta_{bc} \delta_{\nu_1\nu_4}\delta_{\nu_2\nu_3}$
to make the expressions shorter.\\
Inserting the $R_i$ terms from Eqs.~(\ref{eq:r1})-(\ref{eq:r4}) into \ceq{eq:2pgf-freq-1}
and regrouping the terms
leads to the following expression for the symmetric improved two-particle Green's function:
\begin{widetext}
\begin{align}
  \label{eq:2pgf-freq-2}
  G_{abcd}^{\nu_1\nu_2\nu_3\nu_4} &= 
  \big(\delta_{12} - \delta_{14}\big)G_{a}^{\nu_1}G_{c}^{\nu_3}
      -\frac{1}{\beta} \mathcal{G}_a^{\nu_1} \mathcal{G}_b^{\nu_2} \mathcal{G}_c^{\nu_3} \mathcal{G}_d^{\nu_4}   \mathfrak{F}_{abcd}^{\nuall} \; \;  \mbox{ with }\notag\\
  \mathfrak{F}_{abcd}^{\nuall} &=
        \beta \big(\delta_{12} - \delta_{14}\big) 
          \Big(2\sum_i U_{[ai][ai]} n_i + \vartheta_{a}^{\nu_1}\Big)
          \Big(2\sum_j U_{[cj][cj]} n_j + \vartheta_{c}^{\nu_3}\Big) \notag\\
        &+ 2 U_{[ac][bd]}
        + 2 \sum_{i} \Big( U_{[ic][bd]}\xi_{ai}^{\nu_1} + U_{[ac][id]}\xi_{bi}^{\nu_2}  
            + U_{[ai][bd]}\xi_{ci}^{\nu_3} + U_{[ac][bi]}\xi_{di}^{\nu_4}\Big) \notag\\
        &\underbrace{- 4 \phi_{abcd}^{\nu_1-\nu_2} \!+\! 2 f_{abcd}^{\nu_1(\nu_1-\nu_2)} \!+\! 2 f_{cdab}^{\nu_3(\nu_2-\nu_1)}}_{(i)}
        \underbrace{+ 4 \phi_{adcb}^{\nu_1-\nu_4} \!-\! 2 f_{adcb}^{\nu_1(\nu_1-\nu_4)} \!-\! 2 f_{cbad}^{\nu_3(\nu_4-\nu_1)}}_{(ii)}
        \underbrace{- \psi_{abcd}^{\nu_1+\nu_3} \!-\! g_{abcd}^{\nu_1\nu_3} \!-\! g_{dcba}^{\nu_4\nu_2}}_{(iii)}
        + h_{abcd}^{\nu_1\nu_2\nu_3\nu_4} \;.
\end{align}
\end{widetext}
(A pictorial representation of this formula is found in \cfg{fig:gconn-diagrams}.)
The quantities entering here are the antisymmetrized Coulomb interaction  $U$, 
the one-particle density $n_i$ and the correlators  $\vartheta$ and $\xi$ 
which already appeared for the one-particle Green's function in Section~\ref{sec:1pEst}. 
In order to shorten the expression, we singled out the disconnected part
$(\delta_{12}-\delta_{14})G_a^{\nu_1}G_c^{\nu_3}$
of the two-particle Green's function by applying \ceq{eq:sie-1p}.
However, there remains another term proportional to $(\delta_{12}-\delta_{14})$
in $\mathfrak{F}$.
This is rooted in the expansion via equations of motion, which always leads to expressions
involving non-interacting Green's functions $\cal G$.
The 10 terms in the last line of Eq.~(\ref{eq:2pgf-freq-2}) are genuinely related to the two-particle Green's function. 
Further, the frequency structure of the terms $(i)$-$(iii)$ 
resembles the contributions from the particle-hole, transversal particle-hole 
and particle-particle channel \cite{Rohringer_RMP}.
The last term $h_{abcd}^{\nu_1\nu_2\nu_3\nu_4}$ must hence include fully irreducible contributions.

\paragraph*{Two-particle vertex.}
As for two-particle quantities, one is often interested in vertex functions
instead of the Green's function itself, e.g.~when calculating susceptibilities \cite{Georges,Kunes}
or for diagrammatic extensions of DMFT \cite{Toschi,Rubtsov2008,Rohringer_RMP}.
The full vertex $F$ is related to the two-particle Green's function by
\begin{equation}
  \label{eq:def-vertex}
  G_{abcd}^{\nuall} = (\delta_{12}-\delta_{14}) G_{a}^{\nu_1} G_{c}^{\nu_3} 
  - \frac{1}{\beta} G_a^{\nu_1} G_b^{\nu_2} G_c^{\nu_3} G_d^{\nu_4} F_{abcd}^{\nuall}.
\end{equation}
 This bears  a certain similarity to \ceq{eq:2pgf-freq-2}, and 
it  becomes apparent, that the full vertex is given by
\begin{equation}
  \label{eq:vertex}
  F_{abcd}^{\nuall} = \frac{\mathcal{G}_a^{\nu_1} \mathcal{G}_b^{\nu_2} \mathcal{G}_c^{\nu_3} \mathcal{G}_d^{\nu_4}}
    {G_a^{\nu_1} G_b^{\nu_2} G_c^{\nu_3} G_d^{\nu_4}} 
    \mathfrak{F}_{abcd}^{\nuall} \; .
\end{equation}

\paragraph*{Error propagation.}
Assuming that quantities measured in CT-QMC have approximately the same noise amplitude over the whole frequency range,
we can conclude that this holds for $\mathfrak{F}$ as well. We furthermore conclude from Eq.~(\ref{eq:vertex}) that the noise amplitude
is rescaled by $\mathcal{G}^\nu/G^\nu$ in every frequency variable. Since this ratio goes to 1 in the limit of
high frequency, we may finally conclude that the noise amplitude of the vertex $F$ in the high-frequency region is identical
to the noise amplitude of $\mathfrak{F}$ and thus directly proportional to the error of the Monte Carlo simulation. We  hence expect (and will confirm this later) a constant noise level for large Matsubara frequencies.

 In contrast, for conventional CT-QMC calculations in the hybridization expansion  a strong increase of noise in $F$ with increasing Matsubara frequencies is observed. This yields a too noisy vertex at high frequencies so that approaches to circumvent the calculation of $F$ at large frequencies have been developed, as e.g.\ replacing the vertex by its high frequency asymptotics  \cite{Kaufmann}. 
This high noise level of conventional CT-HYB calculations can be understood from Eq.~(\ref{eq:def-vertex}). If we assume a constant noise level of the two-particle Green's function on the left hand side, extracting $F$ by dividing through four Green's functions increases the error four times by a factor $\sim \nu_i$ for large $\nu_i$.

\subsection{Three-leg vertex}\label{sec:threeleg}

Let us further  define the particle-hole three-leg Green's function as
\begin{align}
  \label{eq:threeleg-gf}
  G_{abcd}^{\nu_1\nu_2} = \frac{1}{\beta} & \int_0^\beta d\tau_1 d\tau_2 d\tau_3 
  e^{i\nu_1(\tau_1-\tau_3) - i\nu_2(\tau_2-\tau_3)} \notag \\
  & \times \langle T_\tau d_a^{\phantom{\dag}}(\tau_1) d_b^\dag(\tau_2) [d_c^{\phantom{\dag}} d_d^\dag](\tau_3) \rangle,
\end{align}
where $\nu_1$ and $\nu_2$ are fermionic Matsubara frequencies.
Differentiation with respect to $\tau_1$ and $\tau_2$ 
leads to the symmetric improved version 
\begin{align}
  \label{eq:threeleg-gf-improved}
  G^{\nu_1\nu_2}_{abcd} = &-\beta\delta_{12} (1-n_c) G_a^{\nu_1}
  -\delta_{14} G^{\nu_1}_a G^{\nu_2}_b\notag \\
  &- \mathcal{G}_a^{\nu_1} \mathcal{G}_b^{\nu_2} \left[
    \beta\delta_{12} n_c (2(Un)_a+\vartheta_a^{\nu_1})\right. \notag \\
  &\left.-\delta_{14} \xi_a^{\nu_1} \xi_b^{\nu_2}
  - 2\hat{\phi}^{\nu_1-\nu_2}_{abcd} + \hat{f}^{\nu_1(\nu_1-\nu_2)}_{abcd}
\right],
\end{align}
with the auxiliary definitions
\begin{align}
  \hat{\phi}^\omega_{abcd} = \frac{1}{\beta} & \int d\tau_1 d\tau_2 e^{i\omega(\tau_1-\tau_2)} \notag\\
                                                & \langle T_\tau \sum_{jk} U_{[aj][bk]} [d_j^\dag d_k](\tau_1) [d_d^\dag d_c](\tau_2) \rangle
\end{align}
and
\begin{align}
  \hat{f}^{\nu\omega}_{abcd} = \frac{1}{\beta} & \int d\tau_1 d\tau_2 d\tau_3 e^{i\nu(\tau_1-\tau_2) + i\omega(\tau_2-\tau_3)} \notag\\
                                               & \langle T_\tau q_a(\tau_1) q_b^\dag(\tau_2) [d_d^\dag d_c](\tau_3) \rangle.
\end{align}
Since the $R_2$ term of the symmetric improved estimator of the two-particle Green's function
essentially contains a threeleg Green's function, the derivation of the threeleg symmetric improved estimator
is completely analogous to the one shown in the last subsection of Appendix \ref{app:2p}.

For diagrammatic extensions of DMFT, the associated  three-leg vertex functions are often particularly interesting, 
see e.g. Refs.~\cite{Katanin2009,Rubtsov12,Ayral,Galler2016,vanLoon2018}.
These are related to the three-leg Green's function by subtraction of disconnected
parts and division by two one-particle Green's functions. 
Representative for the variety of definitions, we show the so-called Kernel-2 function in the particle-hole channel (for a definition see \cite{Li2016,Wentzell, Kaufmann}), which can be obtained through the following symmetric improved estimator
\begin{align}
  \label{eq:k2-improved}
  &K^{(2), ph, \nu_1\nu_2}_{abcd}
  \!= \! \left( 4\beta\delta_{12}(Un)_a(Un)_c  \!-  \!4\phi^{\nu_1-\nu_2}_{abcd} \right)
  \left( \! \frac{\mathcal{G}_a^{\nu_1} \mathcal{G}_b^{\nu_2}}{G_a^{\nu_1} G_b^{\nu_2}}  \!-  \!1  \!\right)\notag\\
  &+ 2 \frac{\mathcal{G}_a^{\nu_1} \mathcal{G}_b^{\nu_2}}{G_a^{\nu_1} G_b^{\nu_2}} \big[
    \beta \delta_{12}\vartheta_a^{\nu_1} (Un)_c - \xi_a^{\nu_1} \xi_b^{\nu_2} U_{[cb][da]} + f^{\nu_1(\nu_1-\nu_2)}_{abcd}
\big].
\end{align}

\section{Implementation} \label{sec:Implementation}

As discussed before, the expressions for e.g.\ the one- or two-particle Green's functions derived using the EOM (\ceq{eq:sie-1p} and \ceq{eq:2pgf-freq-2}) can be employed in a CT-HYB simulation to obtain results with asymptotically smaller error for high frequencies. In the simulation, we need to get the QMC estimates for  the values of each individual term contributing to the symmetric improved estimators (cf.\ Tab.~\ref{tab:estimators}), i.e.\ of correlation functions consisting of up to 12 operators with up to 4 different imaginary time or 3 different Matsubara frequency arguments with several components corresponding to possible combinations of the discrete quantum numbers of the operators. We accomplish this by performing worm sampling in our CT-HYB \textsc{w2dynamics}\cite{Wallerberger19} program package.

A full introduction to the CT-HYB algorithm is given in Ref.~\onlinecite{Gull}, 
but for the sake of understanding let us briefly recall some main aspects. 
The starting point is the AIM Hamiltonian Eq.~\eqref{eq:aimham}. 
After integrating out the bath degrees of freedom one can evaluate the thermodynamic partition function 
$Z = \operatorname{Tr} \exp(-\beta H)$ for the impurity by summing over all impurity field configurations. 
The CT-HYB choice of expanding the exponential in the hybridization turns it into a series of ``local'' traces.
These consist of pairs of impurity operators that evolve in imaginary time according 
to the local part of the Hamiltonian. The hybridization with the bath is described by 
a determinant of a matrix that contains, order by order, the hybridization function 
$\Delta$ where the impurity operators in the local trace are.

Each combination of expansion order, imaginary times, orbitals and spins of the local operators describes one point in the space of partition function configurations $\mathcal{C}_Z$. The quantity of interest, e.g.\ a Green's function $G_{ab} = \operatorname{Tr} T_\tau\exp(-\beta H)d_a(\tau_a) d^\dag_b(\tau_b)/Z$, can be obtained from $\mathcal{C}_Z$ either by manipulating a $\mathcal{C}_Z$ configuration accordingly in the measurement step or by directly sampling $\mathcal{C}_{G_{ab}}$ configurations. These are like the $\mathcal{C}_Z$ configurations, but explicitly contain (in this case) two additional operators (``worm'') that appear in the definition of $G_{ab}$. The worm algorithm consists in sampling both $\mathcal{C}_{G_{ab}}$ and $\mathcal{C}_Z$ (for normalization) in one simulation. More details on worm sampling can be found in Ref.~\onlinecite{Gunacker}.

This kind of worm sampling is employed for the calculation of each component $k_{a_1 \dots a_n}$ 
of each correlation function $k$ needed for the symmetric improved estimator (see Table \ref{tab:estimators}). 
When we perform a sampling run in the extended configuration space $\mathcal{C}_Z \oplus \mathcal{C}_{k_{a_1 \dots a_n}}$ which includes partition function configurations and worm configurations for the specific correlation function, the measurement procedure itself trivially consists of counting samples.

As we have seen, the weight in the worm spaces differs from the weight of a similar $Z$ space configuration in the value of the local trace, but we also explicitly add a suitably chosen weight factor $\eta_{k_{a_1 \dots a_n}}$ to balance the number of steps spent in the current worm space and the partition function space. It is not necessary to perform separate runs per component and quantity, but this also makes it simpler in practice to select the appropriate amount of measurements for the desired target error of the final result (to which all components of all quantities may contribute differently).

In both configuration spaces, we allow all moves that only change operators 
connected to hybridization events, i.e.\ in our case pair insertions,
pair removals and some global moves. Additionally, worm insertion and removal steps 
must be employed to change between the two subspaces $\mathcal{C}_Z$ and $\mathcal{C}_{k_{a_1\ldots a_n}}$. 
For estimators with density-like parts only, such as $\phi_{aabb}^{\omega}$, 
this should even in practice be enough to ensure ergodic sampling. For other estimators,
possible quantum number violations (i.e.\ configurations with two sequential operators
that are zero because of commutation relations) and changes in the energies of states
occurring in the local time evolution strongly suppress insertions with large time 
differences between ``compensating'' operators (cf.\ Ref.~\onlinecite{Gunacker,Shinaoka2014}). 
If the worm operators' positions could only be changed in the worm insertion step, 
this would lead to problematically bad statistics for large distances (towards $\Delta \tau = \beta/2$).

\begin{figure}
  \includegraphics[width=\linewidth]{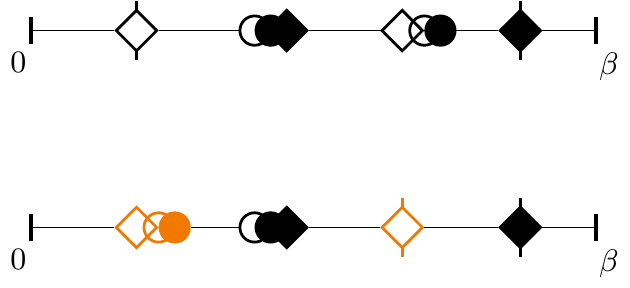}
  \caption{\label{fig:qreplacement}Symbolic representation of a
    ``replacement'' move applied to a worm operator that is at equal
    time with two other operators. Operators are represented as
    symbols (flavors) on the imaginary time axis before (top) and
    after (bottom) the move, with filled symbols representing creators
    and little vertical tags at the symbols' top and bottom signifying
    operators connected to hybridization events. Changes due to the
    move are marked in orange.}
\end{figure}

Therefore, we introduce further moves that shift or replace some 
of the estimator's worm operators analogous to the worm replacement moves\cite{Gunacker}. 
In Ref.~\onlinecite{Gunacker}, these moves transfer the ``worm status'' 
from a worm operator to a hybridization operator, i.e.\ they change which operators 
are connected with hybridization events (and accordingly only change the bath weight). 
Since, contrary to Ref.~\onlinecite{Gunacker}, our estimators contain several operators at equal times, 
the procedure needs to be slightly modified as compared to a simple replacement:
After we select one impurity operator connected to a hybridization event at $\tau_h$
and one worm operator at $\tau_w$ for a replacement, we not only ``exchange'' 
them (changing only the bath part of the weight), but also move any other worm operators
at $\tau_w$ to $\tau_h$ to reconstruct the same equal-time object at another position 
(cf. Fig.~\ref{fig:qreplacement}). This may cause a lower acceptance rate 
compared to simpler replacements, but especially if the other worm operators
are density-like (which may e.g.\ be the case when performed on a $q$), 
these moves are reasonably effective.

\section{Validation}\label{sec:Validation}
The best way to validate CT-QMC results is benchmarking against exact diagonalization (ED). 
To this end CT-QMC results were calculated in \textsc{w2dynamics}\cite{Wallerberger19} and, at the same time,
all estimators were calculated 
exactly by evaluating their Lehmann representation for a small 
Hamiltonian, i.e.\ an impurity model with one orbital and a discrete bath. 
The ED results were also used to confirm the validity of \ceq{eq:2pgf-freq-2}.\\
Specifically, we choose  a bath with
one energy level $\varepsilon_\mathrm{bath} = 0.5$ and hybridization amplitude 
$V = 0.3$. The chemical potential is set to $\mu=-0.1$, the inverse temperature is
chosen to be $\beta=10$ and the local interaction $U=2$. 
At these parameters, every spin-orbital is on average occupied by 0.307 electrons.

\subsection{Self-energy}
The quantities $G^\nu$, $\xi^\nu$ and $\theta^\nu$ were evaluated by performing 1.44$\times 10^9$ QMC measurements on their respective estimators. 
Subsequently the one-particle Green's function and the self-energy were calculated from \ceq{eq:ie-1p} and \ceq{eq:sie-1p}.
\cfg{fig:s-improved} shows a comparison of the improved and the symmetric improved self-energy
as well as the one obtained directly from the CT-HYB Green's function as calculated by worm sampling without improved estimators.
We note that both ``improved'' ways to calculate the self-energy
suffer from lower precision in the low-frequency regime. However, except for the first few Matsubara frequencies, 
where precise quantities can be obtained by conventional $Z$-sampling of the Green's function, 
the symmetric improved one-particle estimator yields considerably better results. 
As anticipated in Section \ref{sec:propagation}, we also observe a better scaling of the error at high Matsubara frequencies.
\begin{figure}
  \includegraphics[width=0.4\textwidth]{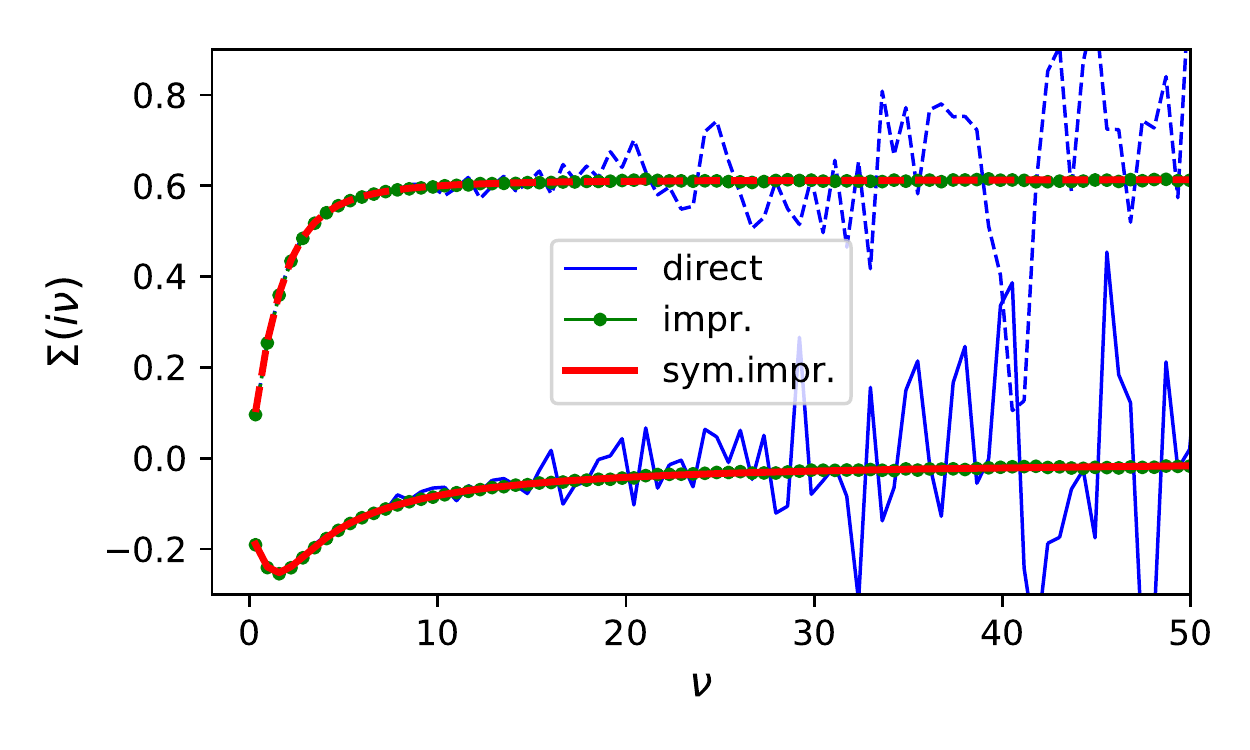}\\
  \includegraphics[width=0.4\textwidth]{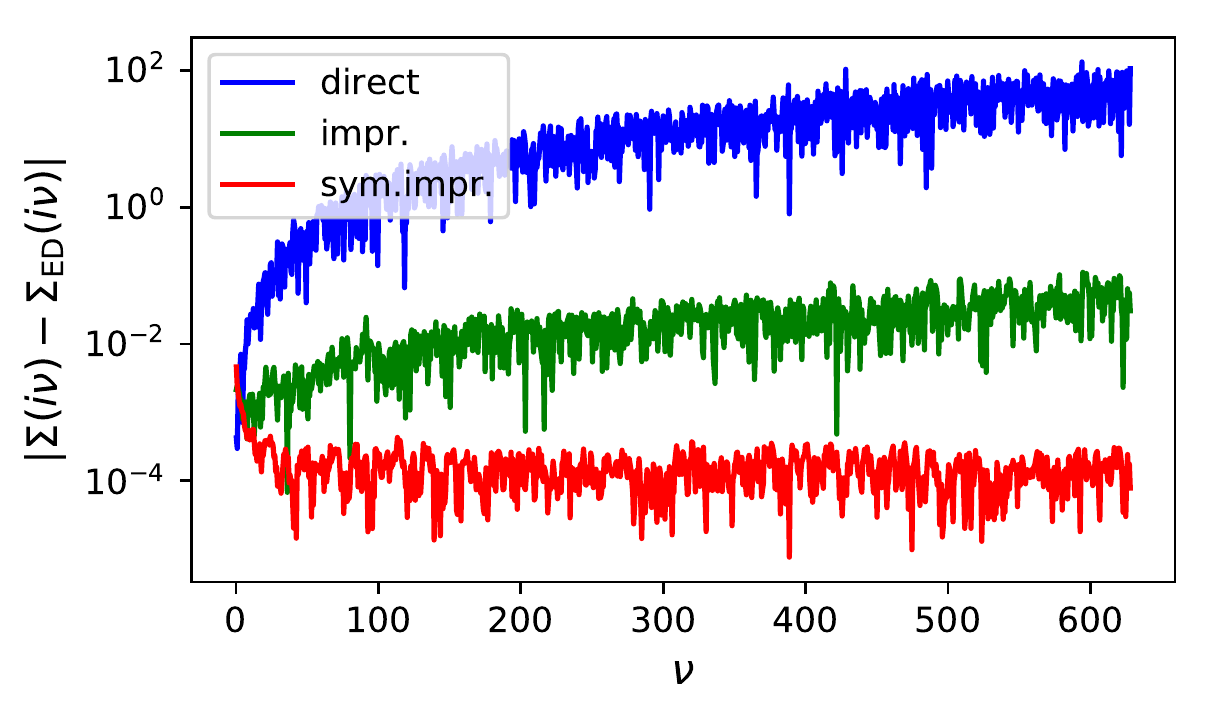}
  \caption{\label{fig:s-improved} Upper panel: Comparison of the self-energy 
    (for the AIM specified in Sec.~\ref{sec:Validation}),
    as calculated from the
    directly measured one-particle Green's function (\emph{direct}) 
    vs.~the result obtained with improved (\emph{impr.}) 
    and symmetrically improved (\emph{sym. impr.}) estimators. Lower panel:
    logplot of the absolute difference to exact diagonalization data. }
\end{figure}
\subsection{Vertex functions}
Vertex functions are related to the two-particle Green's function. In order to assemble the symmetric improved form
by \ceq{eq:2pgf-freq-2}, one needs to sample all seven occurring estimators.
The full vertex $F^{\nu\nu'\omega}\equiv F^{\nu(\nu-\omega)(\nu'-\omega)\nu'}$ can be obtained
from the connected part of the two-particle Green's function 
by ``amputation'' of its legs, i.~e.~division by a product of four Green's functions, cf.~\ceq{eq:def-vertex}.
In the high-frequency case this leads to massive noise amplification, if the two-particle Green's function
is directly computed in QMC.
However, as discussed in Sec.~\ref{sec:2pgf}, this is healed by symmetric improved estimators.
In \cfg{fig:f-improved} we show slices through $F^{\nu\nu'\omega}_{\uparrow\uparrow\downarrow\downarrow}$
at two fixed bosonic frequencies $\omega$. 
In analogy to the one-particle estimator, we get precise results over the whole frequency range,
and in particular, there is no increase of noise at high Matsubara frequencies.
\begin{figure*}
\begin{minipage}{.67\textwidth}
  \includegraphics[width=0.95\textwidth]{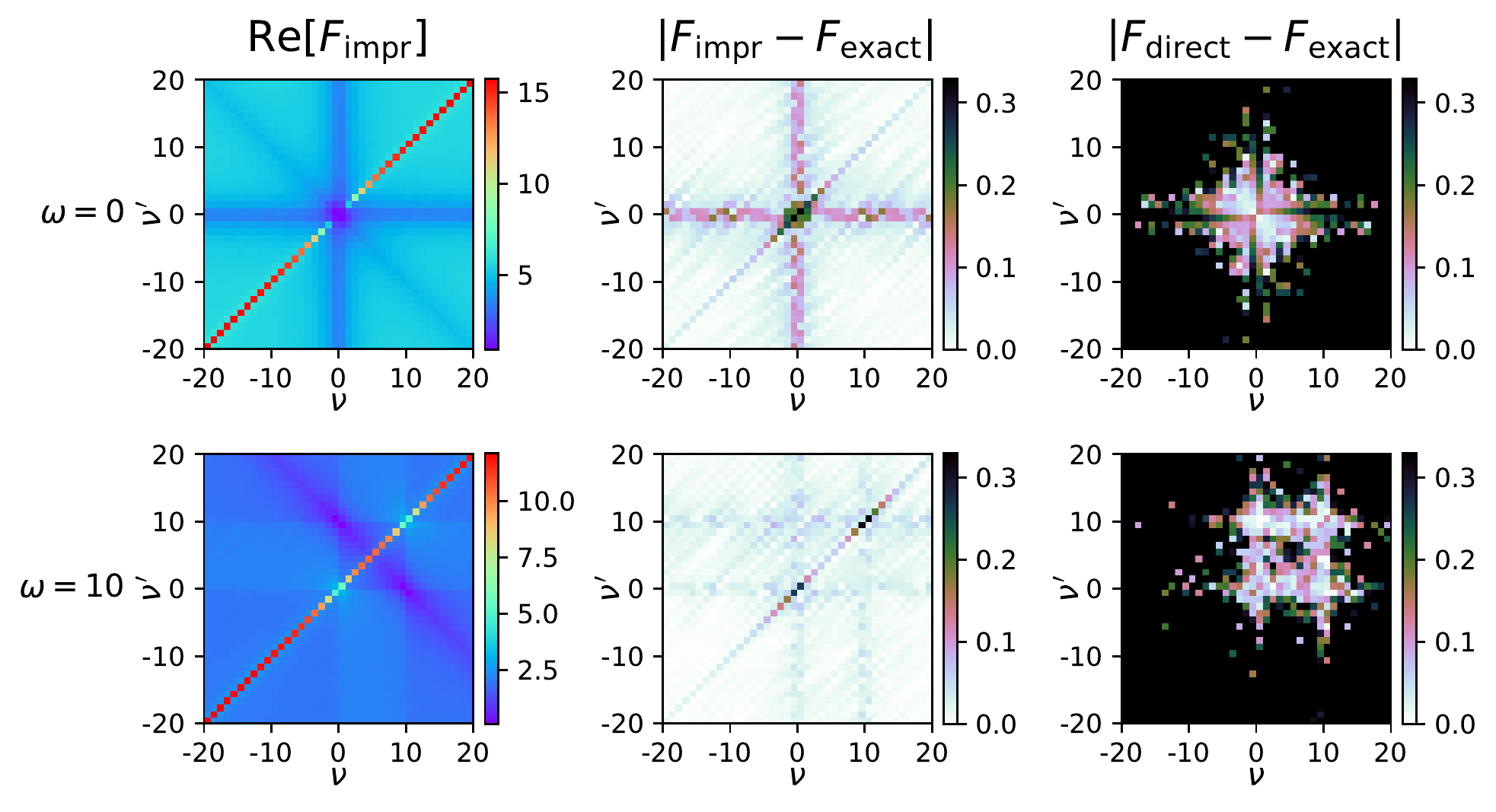}
\end{minipage}\hfill
\begin{minipage}{.31\textwidth}
    \caption{\label{fig:f-improved} Reducible vertex $F^{\nu\nu'\omega}_{\uparrow\uparrow\downarrow\downarrow}$. 
    Upper row: bosonic frequency $\nu_1\!\!-\!\!\nu_2\!=\!\omega\!=\!0$; 
    lower row: $\nu_1\!\!-\!\!\nu_2\! =\! \omega\! =\! 10 \times 2\pi/\beta$.
    First column: $F$ as calculated with symmetric improved estimators; 
    second column: difference of the symmetric improved to the exact result, 
    third column: difference of the conventional calculation (directly measured two-particle Green's function) to the exact result.}
\end{minipage}
\end{figure*}

\subsection{Three-leg vertex}
\ceq{eq:k2-improved} allows us to compute the kernel-2 function $K^{(2),ph,\nu(\nu-\omega)}$
from QMC-estimators that were also used for the full two-particle vertex.
In order to judge the improvement introduced by \ceq{eq:k2-improved}, we 
compute $K^{(2),ph,\nu(\nu-\omega)}_{\uparrow\uparrow\downarrow\downarrow}$ 
not only in this new way, but also
in the way of Ref.~\onlinecite{Kaufmann}
from the three-leg Green's function \ceq{eq:threeleg-gf} measured in QMC.
In \cfg{fig:k2-improved} we compare the result obtained by symmetric
improved estimators to the exact result and to the result of the conventional calculation.
Notably also here the increase of noise at high Matsubara frequencies is absent.

\begin{figure*}
\begin{minipage}{.67\textwidth}
  \includegraphics[width=\textwidth]{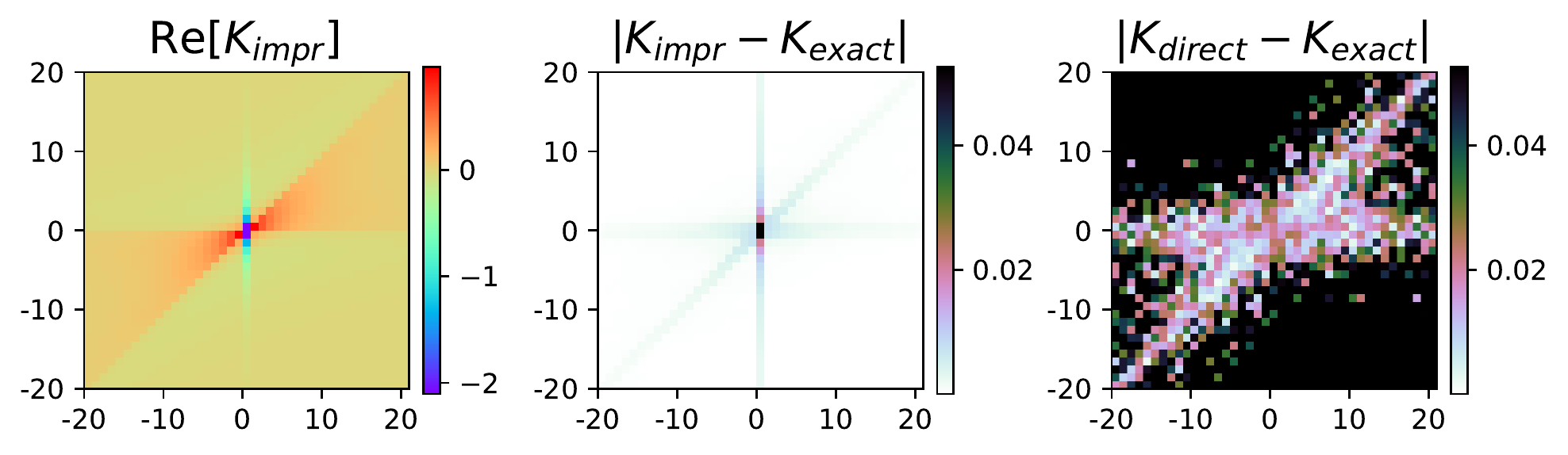}
\end{minipage} \hfill
\begin{minipage}{.31\textwidth}
  \caption{\label{fig:k2-improved} Three-leg kernel function $K^{(2),ph,\nu(\nu-\omega)}_{\uparrow\uparrow\downarrow\downarrow}$, 
    calculated by \ceq{eq:k2-improved} (left), its difference to the exact result (middle), 
    and the difference of the conventional, direct calculation to the exact result (right).}
\end{minipage}
\end{figure*}

\section{Conclusion} \label{sec:Conclusion}
We derived equations of motion for the one- and two-particle Green's function differentiating symmetrically with respect to all
 time arguments. With these symmetric improved estimators, we found a way to compute self-energy and vertex functions in CT-HYB
without suffering from noise that strongly increases at high Matsubara frequencies as in conventional CT-HYB calculations. In fact, our results rapidly
converge towards the exact results, with the exception of the lowest few Matsubara frequencies, where our estimators do not lead to an improvement.
For very weak hybridization the self-energy and vertex on the lowest few Matsubara frequencies can actually be calculated with higher accuracy if no improved
estimators are used.
We hence conclude that in some cases it will be best to combine conventional and improved estimators, using the former for small frequencies and the latter at large frequencies.

\acknowledgments
We thank Markus Wallerberger, Andreas Hausoel, Tin Ribic, Oleg Janson, and Dominique Geffroy for valuable discussions.
This work has been supported  by the  Vienna Scientific Cluster (VSC) Research Center 
funded by the Austrian Federal Ministry of Science, Research and Economy (bmwfw),
the Deutsche Forschungs Gemeinschaft (DFG) through research unit FOR 1346, the 
Austrian Science Fund (FWF) through  project P 30997-N32, 
and the European Research Council under the European Union's Seventh Framework Programme (FP/2007-2013)/ERC grant agreement n. 306447 (AbinitioD$\Gamma$A).  
J.K.~was partially supported by the Leibniz society through the Leibniz competition.
A.K. and G.S. have been supported by the DFG (through SFB 1170 ``ToCoTronics'').
G. S. further acknowledges financial support from the DFG through the W\"urzburg-Dresden 
Cluster of Excellence on Complexity and Topology in Quantum Matter -- \textit{ct.qmat} (EXC 2147, project-id 39085490).
Some of the computational results were obtained using the Vienna Scientific Cluster (VSC).
The authors also gratefully acknowledge the Gauss Centre for Supercomputing e.V. (www.gauss-centre.eu)
for funding this project by providing computing time on the GCS Supercomputer SuperMUC at Leibniz Supercomputing Centre (www.lrz.de).

\appendix

\section{One-particle Symmetric Improved Estimator}\label{app:1P_sym}

The first-order equations (i.e. improved estimators) were already derived in Ref.~\onlinecite{Hafermann}. We nevertheless give a detailed derivation to introduce notation and concepts necessary to derive higher-order estimators (i.e. symmetric improved estimators) in the following. This further sets the stage for the two-particle symmetric improved estimators. 

The time derivative of the one-particle Green's function follows as: 
\begin{align}
\dtau{1} G_a (\tau_1, \tau_2) &= - \dtau{1} \langle T_\tau d_a (\tau_1) \ddag_a (\tau_2) \rangle  \\
&= -\dtau{1} \big( \theta(\tau_1 - \tau_2) \langle d_a (\tau_1) \ddag_a (\tau_2) \rangle -\notag \\
&\phantom{= -\dtau{1} \big(a} \theta(\tau_2-\tau_1) \langle  \ddag_a (\tau_2) d_a (\tau_1) \rangle \big) \\
&= -\delta(\tau_1 - \tau_2) \langle \underbrace{\{d_a, \ddag_a\}}_{=1}  (\tau_1) \rangle  \notag\\
&\phantom{=}-\langle T_\tau (\dtau{1} d_a (\tau_1)) \ddag_a (\tau_2) \rangle  \\
&= -\delta(\tau_1 - \tau_2) - \langle T_\tau [H_{\mathrm{AIM}},d_a] (\tau_1)  \ddag_a (\tau_2) \rangle .
\end{align}

The commutator between the local Hamiltonian and the impurity-annihilation operator on the right hand side can be  calculated as:
\begin{align}
\label{eq:com_H_a}
[H_{\mathrm{AIM}},d_a] &= \frac{1}{2} \sum_{ijkl} U_{ijkl}  (\ddag_i \ddag_j d_l d_k d_a - d_a \ddag_i \ddag_j d_l d_k) \notag \\
&{\phantom{aaa}} + \sum_i \varepstilde_i (\ddag_i d_i d_a - d_a \ddag_i d_i)   \notag \\
&{\phantom{aaa}} + \sum_{Ki}  (V_K^{ii})^* ( \ddag_i c_{Ki} d_a  - d_a \ddag_i c_{Ki}) \\
&= \frac{1}{2} \sum_{ijkl} U_{ijkl}  (\ddag_i \delta_{aj} d_l d_k - \delta_{ai} \ddag_j d_l d_k)  \notag \\
&{\phantom{aaa}} -  \sum_i \varepstilde_i \delta_{ai} d_i   - \sum_{Ki}  (V_K^{ii})^*  \delta_{ai} c_{Ki}  \\
&= -\sum_{jkl} U_{[aj]kl}  \ddag_j d_l d_k  -  \varepstilde_a d_a   - \sum_{K}  (V_K^{aa})^*  c_{Ka},
\end{align}
with hybridization $V_K^{aa}$ and the anti-symmetrizations of the interaction matrix:
\begin{align}
  \label{eq:u-asy}
&\frac{1}{2} \left( U_{ijkl} - U_{jikl} \right) \eqqcolon U_{[ij]kl} \\
 &\frac{1}{2} \left( U_{ijkl} - U_{ijlk} \right) \eqqcolon U_{ij[kl]}.
\end{align}

It is convenient to introduce new operators for the contractions of the interaction matrix with three local operators:
\begin{align}
q^{\phantom{\dag}}_a &= \sum_{jkl} U_{[aj]kl}\ddag_j d_l d_k \\
q^\dagger_{a} &= \sum_{mno} U_{mn[ao]}\ddag_m \ddag_n d_o.
\end{align}

With these equal-time three-particle operators, the one-particle Green's function follows as:
\begin{align}
\label{eq:1p_mixed}
\dtau{1} G_a (\tau_1, \tau_2) &= -\delta(\tau_1 - \tau_2) + \langle T_\tau q_a (\tau_1)  \ddag_a (\tau_2) \rangle \notag\\ 
&{\phantom{aaa}} + \langle T_\tau \varepstilde_a d_a (\tau_1)  \ddag_a (\tau_2) \rangle  \notag\\
&{\phantom{aaa}} + \langle T_\tau \sum_{K}  (V_K^{aa})^*  c_{Ka} (\tau_1)  \ddag_a (\tau_2) \rangle ,
\end{align}
which is the equation of motion for the impurity Green's function of the AIM.

The mixed impurity-bath Green's function can be further calculated by applying the above procedure onto the bath operator once again:
\begin{align}
\dtau{1} \langle T_\tau  c_{Ka} (\tau_1)  \ddag_a (\tau_2) \rangle =& \delta(\tau_1-\tau_2) \langle \underbrace{\{c_{Ka},\ddag_a \}}_{=0} (\tau_1) \rangle \notag \\
&+ \langle T_\tau  [H_{\mathrm{AIM}},c_{Ka}] (\tau_1)  \ddag_a (\tau_2) \rangle
\end{align}

Here, the commutator between the local Hamiltonian and the bath-annihilation operator can be calculated as:
\begin{align}
\label{eq:com_H_c}
[H_{\mathrm{AIM}},c_{Ka}] & =  \sum_{K'i} \varepsilon_{K'i} (\cdag_{K'i} c_{K'i} c_{Ka} - c_{Ka} \cdag_{K'i} c_{K'i})  \notag \\ 
& + \sum_{K'i} V_{K'}^{ii} (\cdag_{K'i} d_i c_{Ka} - c_{Ka} \cdag_{K'i} d_i ) \\
& = -\sum_{K'i} \varepsilon_{K'i} \delta_{K K'} \delta_{ai} c_{K'i}  - \sum_{K'i} V_{K'}^{ii} \delta_{K K'} \delta_{ai} d_i.
\end{align}

Thus we can relate the time-derivative of the mixed impurity-bath Green's function to the impurity Green's function:
\begin{align}
\label{eq:bath_c_a}
(\dtau{1} + \varepsilon_{Ka}) \langle T_\tau  c_{Ka} (\tau_1)  \ddag_a (\tau_2) \rangle = -V_{K}^{aa} \langle T_\tau d_a (\tau_1) \ddag_a (\tau_2) \rangle. 
\end{align}

In order to insert this into \ceq{eq:1p_mixed}, we have to Fourier-transform the above expression with respect to $\tau_1$:
\begin{align}
  \label{eq:1p_mixed_2}
  \int_0^\beta \mathrm{d}\tau_1 e^{i\nu_1\tau_1} \dtau{1}\langle T_\tau  c_{Ka} (\tau_1) & \ddag_a (\tau_2) \rangle\notag\\
  = -\int_0^\beta \mathrm{d}\tau_1 e^{i\nu_1\tau_1}  (\varepsilon_{Ka}& \langle T_\tau  c_{Ka} (\tau_1)  \ddag_a (\tau_2) \rangle\\
+ V_{K}^{aa} &\langle T_\tau d_a (\tau_1) \ddag_a (\tau_2) \rangle )\notag
\end{align}
Now the expression on the left-hand side of \ceq{eq:1p_mixed_2} can be simplified by integration by parts 
and we have
\begin{align}
  \label{eq:1p_mixed_3}
  \int_0^\beta \mathrm{d}\tau_1 & e^{i\nu_1\tau_1} \langle T_\tau  c_{Ka} (\tau_1)  \ddag_a (\tau_2) \rangle\notag\\
  =& \frac{V_K^{aa}}{i\nu_1 - \varepsilon_{Ka}} \int_0^\beta \mathrm{d}\tau_1 e^{i\nu_1\tau_1}\langle T_\tau d_a (\tau_1) \ddag_a (\tau_2) \rangle
\end{align}
Applying the same Fourier transform and integration by parts also to \ceq{eq:1p_mixed} allows us to insert the above result
and we get
\begin{align}
  i\nu_1\!\! \int_0^\beta\!\! \mathrm{d}\tau_1 &e^{i\nu_1\tau_1} 
       \langle T_\tau d_a(\tau_1) d_a^\dag(\tau_2) \rangle = -\!\!\int_0^\beta \!\!\mathrm{d}\tau_1 e^{i\nu_1\tau_1} \delta(\tau_1\!-\!\tau_2)\notag \\
    + & \int_0^\beta \mathrm{d}\tau_1 e^{i\nu_1\tau_1} \langle T_\tau q_a(\tau_1) d^\dag_a(\tau_2) \rangle \notag \\
    + & \tilde{\varepsilon} \int_0^\beta \mathrm{d}\tau_1  e^{i\nu_1\tau_1} \langle T_\tau d_a(\tau_1) d_a^\dag(\tau_2) \rangle \notag \\
       + & \underbrace{\sum_{K} \frac{V_K^{aa} (V_K^{aa})^\ast}{i\nu_1 - \varepsilon_{Ka}}}_{\Delta_a^{\nu_1}} \int_0^\beta \mathrm{d}\tau_1  e^{i\nu_1\tau_1} \langle T_\tau d_a(\tau_1) d_a^\dag(\tau_2) \rangle
\end{align}
Now the terms can be regrouped to express the Green's function:
\begin{align}
  \label{eq:ie}
  \underbrace{(i\nu_1\! -\! \tilde{\varepsilon}_a \!-\! \Delta_a^{\nu_1})}_{[\mathcal{G}_a^{\nu_1}]^{-1}} & \int_0^\beta\!\! \mathrm{d}\tau_1 e^{i\nu_1\tau_1} \langle T_\tau d_a(\tau_1) d_a^\dag(\tau_2) \rangle\notag\\
  = - &\!\! \int_0^\beta \!\!\mathrm{d}\tau_1 e^{i\nu_1\tau_1} \left[
    \delta(\tau_1\!-\!\tau_2) - \langle T_\tau q_a(\tau_1) d_a^\dag(\tau_2) \rangle
  \right]
\end{align}
This expression can be Fourier-transformed with respect to $\tau_2$ by applying $\int_0^\beta \mathrm{d}\tau_2 \mathrm{exp}(-i\nu_2\tau_2)$
to both sides of the equation. Considering the definitions of the Green's function and the function $\xi$,
one arrives at \ceq{eq:ie-1p}.

In order to obtain the symmetric improved estimator, the equal-time two-particle Green's function is differentiated with respect to the impurity creation operator at time $\tau_2$, such that:
\begin{align}
\dtau{2} \langle T_\tau q(\tau_1)  \ddag_a (\tau_2) \rangle =& -\delta(\tau_1 - \tau_2) \langle \{ q_a, \ddag_a \} (\tau_1) \rangle \notag \\
&+ \langle T_\tau q_a (\tau_1)  [H_{\mathrm{AIM}},\ddag_a] (\tau_2) \rangle.
\end{align}

The anti-commutator $\{ q_a, \ddag_a \}$ for diagonal hybridization functions follows as:
\begin{align}
\{ q_a, \ddag_a \} &= \langle \sum_{jkl} U_{[aj]kl} (\ddag_j d_l \delta_{ak} - \ddag_j \delta_{al} d_k) (\tau_1) \rangle \\
&= 2 \sum_{j} U_{[aj][aj]} n_j
\end{align}

The commutator between the local Hamiltonian and the creation operator is calculated in analogy to Eq.\eqref{eq:com_H_a} and follows as:
\begin{align}
[H_{\mathrm{AIM}},\ddag_a] &= q^\dagger_a  + \varepstilde_a \ddag_a  + \sum_{K} V_K^{aa} \cdag_{Ka}. 
\end{align}

Thus:
\begin{align}
\label{eq:q_a}
&\dtau{2} \langle T_\tau q_a (\tau_1)  \ddag_a (\tau_2) \rangle = \\
&-2 \delta(\tau_1 - \tau_2)  \sum_{j} U_{[aj][aj]} n_j + \langle T_\tau q_a (\tau_1) q^\dagger_a (\tau_2) \rangle \notag \\ 
&+ \langle T_\tau q_a (\tau_1) \varepstilde_a \ddag_a (\tau_2) \rangle + \langle T_\tau q_a (\tau_1) \sum_{K} V_K^{aa} \cdag_{Ka} (\tau_2) \rangle. \notag
\end{align}

The mixed bath-impurity expectation value is calculated in analogy to Eqs.\eqref{eq:com_H_c}-\eqref{eq:bath_c_a} and follows as:
\begin{align}
\dtau{2} \langle T_\tau q_a (\tau_1) \cdag_{Ka} (\tau_2) \rangle &= \langle T_\tau q_a (\tau_1) [H_{\mathrm{AIM}},\cdag_{Ka}] (\tau_2) \rangle \\
  (\dtau{2} - \varepsilon_{Ka})\langle T_\tau q_a (\tau_1) \cdag_{Ka} (\tau_2) \rangle &= (V_{K}^{aa})^*\langle T_\tau q_a (\tau_1) \ddag_a (\tau_2) \rangle.
\end{align}
This allows one to express the mixed bath-impurity expectation value as an impurity expectation value. 
Again the equation can be made algebraic by Fourier transforming it, but this time with respect to $\tau_2$:
\begin{align}
  \int_0^\beta \mathrm{d}\tau_2 e^{-i\nu_2\tau_2} & \langle T_\tau q_a(\tau_1) c_{Ka}^\dag(\tau_2) \rangle\notag\\
  = & \frac{(V_{K}^{aa})^\ast}{i\nu_2 - \varepsilon_{Ka}} \int_0^\beta \mathrm{d}\tau_2 e^{-i\nu_2\tau_2} \langle T_\tau q_a(\tau_1) d_a^\dag(\tau_2) \rangle
\end{align}
Re-inserting into the (Fourier-transformed) Eq.\eqref{eq:q_a} gives:
\begin{align}
  i\nu_2 \int_0^\beta \mathrm{d}\tau_2 & e^{-i\nu_2\tau_2} \langle T_\tau q_a(\tau_1) d_a^\dag(\tau_2) \rangle\notag\\ 
   = - &\int_0^\beta \mathrm{d}\tau_2 e^{-i\nu_2\tau_2} \delta(\tau_1 - \tau_2)  \sum_{j} U_{[aj][aj]} n_j\notag\\
  + &\int_0^\beta \mathrm{d}\tau_2 e^{-i\nu_2\tau_2}\langle T_\tau q_a(\tau_1) q_a^\dag(\tau_2) \rangle \notag\\
  + \tilde{\varepsilon_a} &\int_0^\beta \mathrm{d}\tau_2 e^{-i\nu_2\tau_2}\langle T_\tau q_a(\tau_1) d_a^\dag(\tau_2) \rangle \notag\\
  + \underbrace{\sum_{K} \frac{V_K^{aa}(V_K^{aa})^\ast}{i\nu_2 - \varepsilon_{Ka}}}_{\Delta_a^{\nu_2}}&\int_0^\beta \mathrm{d}\tau_2 e^{-i\nu_2\tau_2}\langle T_\tau q_a(\tau_1) d_a^\dag(\tau_2) \rangle \notag
\end{align}
Rearranging gives:
\begin{align}
  \underbrace{(i\nu_2 - \tilde{\varepsilon}_a  - \Delta_a^{\nu_2})}_{[\mathcal{G}_a^{\nu_2}]^{-1}} &\int_0^\beta \mathrm{d}\tau_2 e^{-i\nu_2\tau_2} \langle T_\tau q_a(\tau_1) d_a^\dag(\tau_2) \rangle \notag\\
  = - & \int_0^\beta \mathrm{d}\tau_2 e^{-i\nu_2\tau_2} \big[ 2  \sum_{j} U_{[aj][aj]} n_j \delta(\tau_1-\tau_2)\notag\\
\quad& -  \langle T_\tau q_a(\tau_1) q_a^\dag(\tau_2) \rangle \big]
\end{align}

Together with the definitions of Table \ref{tab:estimators} we obtain the one-particle symmetric improved estimator \ceq{eq:sie-1p}:
\begin{equation}
G_a^\nu = \mathcal{G}_a^\nu + [\mathcal{G}_a^\nu]^2 \big( 2\sum_{j} U_{[aj][aj]}n_j + \vartheta_a^\nu \big).
\end{equation}

\section{Two-particle Symmetric Improved Estimator}\label{app:2p}

In the following we derive the two-particle symmetric improved estimators. Again the conventional improved estimators were already derived in Ref.~\onlinecite{Hafermann}. 
The procedure in deriving the higher order (up to fourth-order) equations is in principle equivalent to the one-particle symmetric improved estimators. Nevertheless, the equations are more involved due to the necessity of considering multiple hierarchies of equations of motions. Repeating derivations (such as the explicit calculation of mixed impurity-bath expectation values) are omitted.
\subsection*{First Order}
Applying the time derivative onto the first annihilation operator of the two-particle Green's function gives:
\begin{align}
\dtau{1} G&_{abcd}^{\tau_1,\tau_2,\tau_3,\tau_4} =  \notag \\
&R_1 +  \langle T_\tau \underbrace{[H_{\mathrm{AIM}},d_a]}_{\mathclap{\phantom{aaaaaa}-q_a - \varepstilde_a d_a - \sum_K (V_K^{aa})^* c_{Ka}}}(\tau_1) \ddag_b (\tau_2) d_c(\tau_3) \ddag_d (\tau_4) \rangle \\[0.25cm]
(\dtau{1} + &\varepstilde_a ) \ G_{abcd}^{\tau_1,\tau_2,\tau_3,\tau_4} =  \notag\\
&  R_1 -  \underbrace{\langle T_\tau q_a (\tau_1) \ddag_b (\tau_2) d_c(\tau_3) \ddag_d (\tau_4) \rangle}_{\eqqcolon S_1}  \notag \\ 
& - \underbrace{\langle T_\tau \sum_K (V_K^{aa})^* c_{Ka} (\tau_1) \ddag_b (\tau_2) d_c(\tau_3) \ddag_d (\tau_4) \rangle}_{\rightarrow\Delta_a^{\nu_1}\ G_{abcd}^{\nu_1,\tau_2,\tau_3,\tau_4} }  
\end{align}
Again, Fourier transformation with respect to the first time argument converts the differential equation 
to an algebraic one, and we have
\begin{equation}
  G_{abcd}^{\nu_1,\tau_2,\tau_3,\tau_4} = \mathcal{G}_a^{\nu_1}\left( -R_{1,abcd}^{\nu_1,\tau_2,\tau_3,\tau_4} + S_{1,abcd}^{\nu_1,\tau_2,\tau_3,\tau_4}\right).
\label{eq:1ord}
\end{equation}

$R_1$ describes the equal-time contribution due to the derivation of the time-ordering operator:
\begin{align}
  R_{1,abcd}^{\tau_1,\tau_2,\tau_3,\tau_4} = &\delta(\tau_1 - \tau_2) \langle T_{\tau}  \{d_a,\ddag_b\} (\tau_1) d_c(\tau_3) \ddag_d (\tau_4) \rangle + \notag \\
&\delta(\tau_1 - \tau_4) \langle T_{\tau}  \{d_a,\ddag_d\} (\tau_1) \ddag_b(\tau_2) d_c (\tau_3) \rangle \\[0.25cm]
= & \delta(\tau_1 - \tau_2) \delta_{ab} \underbrace{\langle T_{\tau} d_c(\tau_3) \ddag_d (\tau_4) \rangle}_{-G_{cd}^{\tau_3 \tau_4}} \notag \\
- &\delta(\tau_1 - \tau_4) \delta_{ad} \underbrace{\langle  T_{\tau} d_c (\tau_3) \ddag_b(\tau_2)  \rangle}_{-G_{cb}^{\tau_3 \tau_2}} 
\end{align}

At this stage one can already observe that it is possible to apply further (imaginary) time derivatives, or equivalently, hierarchies of equations of motion onto
the $R_1$ term. The corresponding time-derivatives of the one-particle Green's function are given by the one-particle symmetric improved estimators and where already calculated explicitly in Appendix~\ref{app:1P_sym}.

\subsection*{Second Order}
Applying the time derivative onto the first creation operator of $S_1$ gives:
\begin{align}
\dtau{2}  S_1  &= \notag \\
&R_2 +\langle T_\tau q_a (\tau_1)  \underbrace{[H_{\mathrm{AIM}},\ddag_b]}_{\mathclap{q^\dagger_b + \varepstilde_b \ddag_b + \sum_K V_K^{bb} \cdag_{Kb}}} (\tau_2) d_c(\tau_3) \ddag_d (\tau_4) \rangle \\[0.25cm]
(\dtau{2} -& \varepstilde_b ) S_1  = \notag \\
&R_2 + \underbrace{\langle T_\tau q_a (\tau_1)  q^\dagger_b  (\tau_2) d_c(\tau_3) \ddag_d (\tau_4) \rangle}_{\eqqcolon S_2} \notag\\
&+ \underbrace{\langle T_\tau q_a (\tau_1)  \sum_K V_K^{bb} \cdag_{Kb}  (\tau_2) d_c(\tau_3) \ddag_d (\tau_4) \rangle}_{\rightarrow\Delta_b^{\nu_2} S_1^{\tau_1,\nu_2,\tau_3,\tau_4}} \\[0.25cm]
  S_1^{\tau_1,\nu_2,\tau_3,\tau_4} =& \mathcal{G}_b^{\nu_2} \left( R_2^{\tau_1,\nu_2,\tau_3,\tau_4} + S_2^{\tau_1,\nu_2,\tau_3,\tau_4} \right)
\end{align}

Fourier-transforming this also with respect to $\tau_1$ and plugging it into Eq.~\eqref{eq:1ord} yields:
\begin{multline}
  G_{abcd}^{\nu_1,\nu_2,\tau_3,\tau_4} = \mathcal{G}_a^{\nu_1} \left( -R_{1,abcd}^{\nu_1,\nu_2,\tau_3,\tau_4} + \mathcal{G}_b^{\nu_2} \times \right.\\
  \left.\left( R_{2,abcd}^{\nu_1,\nu_2,\tau_3,\tau_4} + S_{2,abcd}^{\nu_1,\nu_2,\tau_3,\tau_4} \right) \right)
\label{eq:2ord}
\end{multline}

$R_2$ describes the equal-time contribution due to the derivation of the time-ordering operator of $S_1$:
\begin{align}
  R_{2,abcd}^{\tau_1,\tau_2,\tau_3,\tau_4} = &-\delta(\tau_1 - \tau_2) \langle T_\tau \underbrace{\{q_a,\ddag_b\}}_{\mathclap{2 \sum_{ef} U_{[ae][bf]} \ddag_e d_f}}(\tau_1)  d_c(\tau_3) \ddag_d (\tau_4) \rangle \notag \\ 
& + \delta(\tau_2 - \tau_3) \langle T_\tau q_a (\tau_1)  \{\ddag_b, d_c\} (\tau_2) \ddag_d (\tau_4) \rangle \\[0.25cm]
= &-2\delta(\tau_1 \! - \! \tau_2) \langle T_\tau \sum_{ef} \!  U_{[ae][bf]} \ddag_e d_f  (\tau_1)  d_c(\tau_3) \ddag_d (\tau_4) \rangle  \notag \\
  &+ \delta(\tau_2 \! - \! \tau_3) \delta_{bc} \langle T_\tau q_a (\tau_1) \ddag_d (\tau_4) \rangle\label{eq:r2-time}
\end{align}
Time derivatives of $R_2$ with respect to $\tau_3$ and $\tau_4$ are shown below.

\subsection*{Third Order}
Applying the time derivative onto the remaining annihilation operator of $S_2$ gives:
\begin{align}
\dtau{3} S_2 &= \notag \\
&R_3 + \langle T_\tau q_a (\tau_1) q^\dagger_b (\tau_2) \underbrace{[H_{\mathrm{AIM}},d_c]}_{\mathclap{-q_c - \varepstilde_c d_c - \sum_K (V_K^{cc})^* c_{Kc}}}(\tau_3) \ddag_d (\tau_4) \rangle \\[0.25cm]
(\dtau{3}+&\varepstilde_c) S_2 = \notag \\
&R_3 - \underbrace{\langle T_\tau q_a (\tau_1) q^\dagger_b (\tau_2) q_c(\tau_3) \ddag_d (\tau_4) \rangle}_{\eqqcolon S_3} \notag\\
&- \underbrace{\langle T_\tau q_a (\tau_1) q^\dagger_b (\tau_2)  \sum_K (V_K^{cc})^* c_{Kc} (\tau_3) \ddag_d (\tau_4) \rangle}_{\rightarrow\Delta_c^{\nu_3} S_2} \\[0.25cm] 
  S_{2,abcd}^{\tau_1,\tau_2,\nu_3,\tau_4} =& \mathcal{G}_c^{\nu_3} \left( -R_{3,abcd}^{\tau_1,\tau_2,\nu_3,\tau_4} + S_{3,abcd}^{\tau_1,\tau_2,\nu_3,\tau_4} \right)
\end{align}

Fourier-transforming this expression for $S_2$ now also with respect to $\tau_1$ and $\tau_2$, 
and plugging it into Eq.~\eqref{eq:2ord} yields:
\begin{multline}
  G_{abcd}^{\nu_1,\nu_2,\nu_3,\tau_4} = \mathcal{G}_a^{\nu_1} \big( -R_{1,abcd}^{\nu_1,\nu_2,\nu_3,\tau_4} + \mathcal{G}_b^{\nu_2} \times \\
  \big( R_{2,abcd}^{\nu_1,\nu_2,\nu_3,\tau_4} + \mathcal{G}_c^{\nu_3} \big( -R_{3,abcd}^{\nu_1,\nu_2,\nu_3,\tau_4} + S_{3,abcd}^{\nu_1,\nu_2,\nu_3,\tau_4} \big) \big) \big)
\label{eq:3ord}
\end{multline}

$R_3$ describes the equal-time contribution due to the derivation of the time-ordering operator of $S_2$:
\begin{align}
  R_{3,abcd}^{\tau_1,\tau_2,\tau_3,\tau_4} &= \delta(\tau_1 - \tau_3) \langle T_\tau \underbrace{\{q_a, d_c\}}_{\mathclap{\sum_{fg}U_{[ac]fg} d_g d_f}} (\tau_1) q^\dagger_b (\tau_2)  \ddag_d (\tau_4) \rangle \notag \\
\qquad- \delta(\tau_2 - \tau_3) & \langle T_\tau q_a (\tau_1) \underbrace{\{q^\dagger_b, d_c\}}_{\mathclap{ 2 \sum_{ij}U_{[ci][bj]} \ddag_i d_j}} (\tau_2) \ddag_d (\tau_4) \rangle \notag \\
\qquad+ \delta(\tau_3 - \tau_4) & \langle T_\tau q_a (\tau_1) q^\dagger_b (\tau_2) \{d_c,\ddag_d\} (\tau_3) \rangle  \\[0.25cm]
\qquad = \delta(\tau_1 - \tau_3) &\langle T_\tau \sum_{fg} U_{[ac]fg} d_g d_f (\tau_1) q^\dagger_b (\tau_2)  \ddag_d (\tau_4) \rangle \notag \\
\qquad  - 2\delta(\tau_2 - \tau_3)& \langle T_\tau q_a (\tau_1) \sum_{ij} U_{[ci][bj]} \ddag_i d_j (\tau_2) \ddag_d (\tau_4) \rangle \notag \\
\qquad  + \delta(\tau_3 - \tau_4) &\delta_{cd} \langle T_\tau q_a (\tau_1) q^\dagger_b (\tau_2) \rangle  
  \label{eq:r3-time}
\end{align}
Time derivatives of $R_3$ with respect to $\tau_4$ are shown below.

\subsection*{Fourth Order}
Applying the time derivative onto the remaining creation operator of $S_3$ gives:
\begin{align}
\dtau{4} S_3 & = \notag \\
&  R_4 + \langle T_\tau q_a (\tau_1) q^\dagger_b (\tau_2) q_c (\tau_3) \underbrace{[H_{\mathrm{AIM}},\ddag_d]}_{\mathclap{q^\dagger_d + \varepstilde_d \ddag_d + \sum_K V_K^{dd} \cdag_{Kd}}} (\tau_4) \rangle \\[0.25cm]
(\dtau{4}-&\varepstilde_d ) S_3 = \notag \\
&  R_4 + \underbrace{\langle T_\tau q_a (\tau_1) q^\dagger_b (\tau_2) q_c (\tau_3) q^\dagger_d (\tau_4) \rangle}_{\eqqcolon S_4} \notag\\
& + \underbrace{\langle T_\tau q_a (\tau_1) q^\dagger_b (\tau_2) q_c (\tau_3) \sum_K V_K^{dd} \cdag_{Kd} (\tau_4) \rangle}_{\rightarrow\Delta_d^{\nu_4} S_3} \\[0.25cm]
  S_{3,abcd}^{\tau_1,\tau_2,\tau_3,\nu_4} =& \mathcal{G}_d^{\nu_4} \left( R_{4,abcd}^{\tau_1,\tau_2,\tau_3,\nu_4} + S_{4,abcd}^{\tau_1,\tau_2,\tau_3,\nu_4} \right) 
\end{align}

Fourier-transforming $S_3$ with respect to the remaining time arguments and inserting it into Eq.~\eqref{eq:3ord} yields:
\begin{multline}
  G_{abcd}^{\nu_1,\nu_2,\nu_3,\nu_4} = \mathcal{G}_a^{\nu_1} \big[ -R_{1,abcd}^{\nu_1,\nu_2,\nu_3,\nu_4} + \mathcal{G}_b^{\nu_2} \times \\
  \big[ R_{2,abcd}^{\nu_1,\nu_2,\nu_3,\nu_4} + \mathcal{G}_c^{\nu_3}\big( -R_{3,abcd}^{\nu_1,\nu_2,\nu_3,\nu_4} + \mathcal{G}_d^{\nu_4} \times\\
\big( R_{4,abcd}^{\nu_1,\nu_2,\nu_3,\nu_4} + S_{4,abcd}^{\nu_1,\nu_2,\nu_3,\nu_4} \big)  \big) \big] \big],
\label{eq:4ord}
\end{multline}
where $S_{4,abcd}^{\nu_1,\nu_2,\nu_3,\nu_4} = h_{abcd}^{\nu_1,\nu_2,\nu_3,\nu_4}/\beta $, as defined in Table \ref{tab:estimators}.
$R_4$ describes the equal-time contribution due to the derivation of the time-ordering operator of $S_3$:
\begin{align}
  \label{eq:2pgf-rec}
  R_{4,abcd}^{\tau_1,\tau_2,\tau_3,\tau_4} = &-\delta(\tau_1 - \tau_4) \langle T_\tau \underbrace{\{q_a,\ddag_d\}}_{\mathclap{2 \sum_{ef} U_{[ae][df]} \ddag_e d_f}}(\tau_1) q^\dagger_b (\tau_2) q_c (\tau_3) \rangle  \notag \\
  +\delta(\tau_2 &- \tau_4) \langle T_\tau q_a (\tau_1) \underbrace{\{q^\dagger_b,\ddag_d\}}_{\mathclap{\sum_{hi} U_{hi[bd]} \ddag_h \ddag_i}} (\tau_2) q_c (\tau_3) \rangle \notag \\
  -\delta(\tau_3 &- \tau_4) \langle T_\tau q_a (\tau_1) q^\dagger_b (\tau_2) \underbrace{\{q_c, \ddag_d\}}_{\mathclap{2 \sum_{kl} U_{[ck][dl]} \ddag_k d_l}} (\tau_3) \rangle \\[0.25cm]
  = -2\delta(\tau_1 &- \tau_4) \langle T_\tau \sum_{ef} U_{[ae][df]} \ddag_e d_f (\tau_1) q^\dagger_b (\tau_2) q_c (\tau_3) \rangle  \notag \\
  +\delta(\tau_2 &- \tau_4) \langle T_\tau q_a (\tau_1) \sum_{hi} U_{hi[bd]} \ddag_h \ddag_i (\tau_2) q_c (\tau_3) \rangle \notag\\
  -2\delta(\tau_3 &- \tau_4) \langle T_\tau q_a (\tau_1) q^\dagger_b (\tau_2) \sum_{kl} U_{[ck][dl]} \ddag_k d_l (\tau_3) \rangle 
\end{align}

Unlike the previous terms $R_1,R_2,R_3$, the above $R_4$ cannot be derived any further.

\subsection*{Third and fourth order of $R_2$ and $R_3$}
Looking at equation \ceq{eq:4ord} one notices that in fact only $R_4$ and $S_4$ are
multiplied by a product of four non-interacting Green's functions. $R_1$ consists of one-particle
Green's functions, which can be expressed by their symmetric improved estimators.
However, $R_2$ and $R_3$ are multiplied only by products of 2 and 3 non-interacting
Green's functions, respectively. This means that we need to express them by their 
equations of motion recursively. \\
First, let us take a closer look at the first term of $R_2$, \ceq{eq:r2-time}.
It is very similar to the three-leg Green's function of \ceq{eq:threeleg-gf},
the only difference being in the operator
ordering and the additional $U$-matrix. The following steps may therefore 
also serve as a derivation of the threeleg improved estimator.\\
Taking the time-derivative with respect to $\tau_3$ generically yields
\begin{align}
  \dtau{3} &\langle T_\tau \sum_{ef} \!  U_{[ae][bf]} \ddag_e d_f  (\tau_1)  d_c(\tau_3) \ddag_d (\tau_4) \rangle\notag\\
   = R_{21} &+ \langle T_\tau \sum_{ef}\! U_{[ae][bf]} \ddag_e d_f (\tau_1) \dtau{3} d_c(\tau_3) d^\dag_d(\tau_4) \rangle
\end{align}
Analogously as before, we express the time derivative by the Heisenberg equation of motion and Fourier-transform
the equation with respect to $\tau_3$ to make it algebraic. 
We thus obtain
\begin{align}
  \label{eq:r2-rec1}
  \int_0^\beta \mathrm{d}\tau_3 & e^{i\nu_3\tau_3} \langle \sum_{ef} \!  U_{[ae][bf]} \ddag_e d_f  (\tau_1)  d_c(\tau_3) \ddag_d (\tau_4) \rangle\notag\\
  = \mathcal{G}_c^{\nu_3} &\int_0^\beta \mathrm{d}\tau_3 e^{i\nu_3\tau_3} \big[ 
     - R_{21}^{\tau_1,\tau_3,\tau_4} \notag\\
    & + \langle \sum_{ef} \!  U_{[ae][bf]} \ddag_e d_f  (\tau_1)  q_c(\tau_3) \ddag_d (\tau_4) \rangle \big]
\end{align}
The rest-term $R_{21}$ that originates from the derivative of the time-ordering operator,
is
\begin{align}
  R_{21}^{\tau_1,\tau_3,\tau_4} = & \delta(\tau_1-\tau_3)
    \langle T_\tau \big[\sum_{ef} U_{[ae][bf]}d_e^\dag d_f, d_c\big](\tau_3) d_d^\dag(\tau_4) \rangle\notag\\
    +& \delta(\tau_3-\tau_4) \langle T_\tau \sum_{ef} U_{[ae][bf]}d_e^\dag d_f \{ d_c, d_d^\dag \}(\tau_3) \rangle.
\end{align}
We want to emphasize that due to the bosonic operator $d^\dag d$, a commutator occurs in the first line.
After evaluation of the commutator and anti-commutator, we get
\begin{align}
  \label{eq:r21}
  R_{21}^{\tau_1,\tau_3,\tau_4} = & \delta(\tau_1-\tau_3) U_{[ac][bd]} \langle T_\tau d_d(\tau_1) d^\dag_d(\tau_4) \rangle\notag\\
  + & \delta(\tau_3-\tau_4) \delta_{cd} \sum_{ef} U_{[ae][bf]} n_{ef}
\end{align}
In order to obtain the final form, we also have to apply the equation of motion with respect
to $\tau_4$. In $R_{21}$ this concerns only the first term. This is, however, just a one-particle
Green's function and we can insert the improved-estimator formula \ceq{eq:ie-1p} and we thus get
\begin{align}
  \label{eq:r21-impr}
  R_{21}^{\tau_1,\tau_3,\nu_4} = &\delta(\tau_1\!-\!\tau_3)  \sum_f U_{[ac][bf]}  \mathcal{G}_d^{\nu_4} \big[ \!-\!e^{-i\nu_4\tau_1}\delta_{fd}\notag\\
  & + \int_0^\beta \mathrm{d}\tau_4 e^{-i\nu_4\tau_4} \langle T_\tau d_f(\tau_1)q_d^\dag(\tau_4) \rangle \big]\notag\\
  + &e^{-i\nu_4\tau_3}  \delta_{cd} \sum_{ef} U_{[ae][bf]} n_{ef}
\end{align}
Whereas the rest-term $R_{21}$ contains only the one-particle improved estimator and the occupation number,
we can differentiate the last term in equation \ceq{eq:r2-rec1} by its last time argument $\tau_4$. 
The same procedure as above yields now
\begin{align}
  \label{eq:r2-rec2}
  \int_0^\beta \mathrm{d}\tau_4 & e^{-i\nu_4\tau_4} \langle \sum_{ef} \!  U_{[ae][bf]} \ddag_e d_f  (\tau_1)  q_c(\tau_3) \ddag_d (\tau_4) \rangle\notag\\
  = \mathcal{G}_d^{\nu_4} &\int_0^\beta \mathrm{d}\tau_4 e^{-i\nu_4\tau_4} \big[ 
     R_{22}^{\tau_1,\tau_3,\tau_4} \notag\\
    & + \langle \sum_{ef} \!  U_{[ae][bf]} \ddag_e d_f  (\tau_1)  q_c(\tau_3) q^\dag_d (\tau_4) \rangle \big]
\end{align}
The rest-term $R_{22}$ again requires some precaution, since it contains both a commutator
and an anti-commutator:
\begin{align}
  R_{22}^{\tau_1,\tau_3,\tau_4} &= \delta(\tau_1-\tau_4) \langle T_\tau  \big[ \sum_{ef} \! U_{[ae][bf]} \ddag_e d_f ,\ddag_d \big](\tau_1) q_c(\tau_3) \rangle\notag\\
  &-\delta(\tau_3-\tau_4) \langle T_\tau \sum_{ef} \! U_{[ae][bf]} \ddag_e d_f (\tau_1) \{ q_c, \ddag_d \}(\tau_3)\rangle
\end{align}
After evaluation of the commutator and the anti-commutator, we obtain
\begin{align}
  \label{eq:r22}
  &R_{22}^{\tau_1,\tau_3,\tau_4} = \delta(\tau_1-\tau_4) \langle T_\tau \sum_e U_{[ae][bd]} d_e^\dag(\tau_1) q_c(\tau_3) \rangle\notag\\
                                  &\quad-2 \delta(\tau_3-\tau_4) \langle \sum_{ef} \!  U_{[ae][bf]} \ddag_e d_f  (\tau_1)
  \sum_{gh} \!  U_{[ag][bh]} \ddag_g d_h  (\tau_3) \rangle.
  \end{align}
We are now able to write down the full expression for $R_2$ by combining \ceq{eq:r2-time} with \ceq{eq:r21-impr} and \ceq{eq:r22}. 
Since in the end we need it in Matsubara frequencies, it is of advantage to perform a Fourier
transform with respect to all time arguments already here. Furthermore this allows us to make
the equation more compact by using the definitions of Table \ref{tab:estimators} and the relation
$\langle T_\tau q_j(\tau) d_k^\dag \rangle = \langle T_\tau d_k(\tau) q_j^\dag \rangle$, such that
we finally arrive at \ceq{eq:r2}.

For $R_3$ we only have to apply the equation of motion with respect to $\tau_4$ to the first two terms of \ceq{eq:r3-time}.
Again we perform a Fourier transform with respect to all time arguments
and compactify the expression by the definitions of Table \ref{tab:estimators}. 
Thus, we obtain \ceq{eq:r3}.

\section{Lehmann representations for two-frequency objects}
For objects depending on a single fermionic/bosonic frequency, the 
standard Lehmann representations for fermionic/bosonic Green's functions can be used. 
For objects depending on two fermionic and one bosonic Green's function, we use
the form published in \cite{Toschi}.
Additionally, we need it for 
\begin{align}
  f^{\nu\omega} = \int_0^\beta & d\tau_1 d\tau_2 d\tau_3 e^{i[\nu(\tau_1-\tau_2) + \omega(\tau_2-\tau_3)]}\notag\\
  &\times\langle \mathcal{T}_\tau F_1(\tau_1) F_2(\tau_2) B(\tau_3) \rangle
\end{align}
Here, $F_i$ ($B$) are fermionic (bosonic) operators, 
and $\nu$ ($\omega$) are fermionic (bosonic) frequencies. 
Inserting the eigenbasis of the Hamiltonian and evaluating the integrals
analytically, we obtain
\begin{align}
  f^{\nu\omega} = \frac{1}{Z} &\sum_{mnl} F_1^{mn}F_2^{nl}B^{lm}
    \frac{1}{-i\nu + i\omega + E_n - E_l}\notag \\
        \times&\bigg[
        \frac{e^{-\beta E_l} - e^{-\beta E_m}}{i\omega + E_m - E_l}
      + \frac{e^{-\beta E_n} + e^{-\beta E_m}}{i\nu + E_m - E_n}
    \bigg]\notag \\
    + \frac{1}{Z} &\sum_{mnl} F_2^{mn} F_1^{nl} B^{lm}
      \frac{1}{i\nu + E_n - E_l}\notag  \\
          \times&\bigg[
          \frac{e^{-\beta E_l} - e^{-\beta E_m}}{i\omega + E_m - E_l}
        + \frac{e^{-\beta E_n} + e^{-\beta E_m}}{-i\nu + i\omega + E_m - E_n}
      \bigg]
\end{align}
Furthermore, we need
\begin{align}
  g^{\nu\omega} = \int_0^\beta & d\tau_1 d\tau_2 d\tau_3  e^{i[\nu_1(\tau_1-\tau_3) + \nu_2(\tau_2-\tau_3)]}\notag\\
  &\times \langle \mathcal{T}_\tau F_1(\tau_1) F_2(\tau_2) B(\tau_3) \rangle.
\end{align}
with two fermionic frequencies $\nu_1$, $\nu_2$. 
The result is
\begin{align}
  g^{\nu\omega} = \frac{1}{Z} &\sum_{mkl} F_1^{mn}F_2^{nl}B^{lm}
    \frac{1}{i\nu_2 + E_n - E_l} \notag\\
      \times&\bigg[
          \frac{e^{-\beta E_l} - e^{-\beta E_m}}{i\nu_1 + i\nu_2 + E_m - E_l}
        + \frac{e^{-\beta E_n} + e^{-\beta E_m}}{i\nu_1 + E_m - E_n}
      \bigg]\notag\\
    + \frac{1}{Z} &\sum_{mnl} F_2^{mn} F_1^{nl} B^{lm}
      \frac{1}{i\nu_1 + E_n - E_l}\notag\\
        \times&\bigg[
          \frac{e^{-\beta E_l} - e^{-\beta E_m}}{i\nu_1 + i\nu_2 + E_m - E_l}
        + \frac{e^{-\beta E_n} + e^{-\beta E_m}}{i\nu_2 + E_m - E_n}
      \bigg]
\end{align}
\newpage
\bibliography{bibliography}
 
\vfill\eject

\end{document}